 \documentclass[12pt]{article}
  \usepackage{epsfig}
\newcommand{\be}{\begin{equation}}
\newcommand{\ee}{\end{equation}}
\newcommand{\ba}{\begin{eqnarray}}
\newcommand{\ea}{\end{eqnarray}}

\newcommand{\inc}{{\it i}}
\newcommand{\smallN}{\mbox{{\tiny $N$}}}

\newcommand{\erbold}{\mbox{{\boldmath $\vec{r}$}}}

\newcommand{\Phibold}{\mbox{{\boldmath $\vec{\Phi}$}}}

\begin{document}
% \date{}
\title{
  ~~~~~~~~~~~~~~~~~~~~~~~~~~~~~~~~~~~~~~~~~~~~~~~~~~~~~~~~~~~~~~~~~~~~~~~~~~~~~~~~~
  ${~}^{\underline{astro-ph/0305344}}$\\
  ${~}^{{Published~in~the}}$~~~~~~~~~~~~~~~~~~~~~~~~~~~~~~~~~~~~~~~~~~\\
  ~~~~~~~~~~~~~${~}^{\underline{Journal~of~Mathematical~Physics,~Vol.~44,~pp.~5958~-~5977~(2003)}}$\\
 ~\\
 {\Large{\bf Gauge Symmetry
of the N-body Problem
\\ in the Hamilton-Jacobi Approach.  }}\\ }
 \author{ {\Large{Michael Efroimsky}}\\ {\small{US Naval Observatory,
 Washington DC 20392 USA}}\\ {\small{e-mail: efroimsk @
 ima.umn.edu~}}\\ ~\\ { {{and}}}\\ ~\\ {\Large{Peter Goldreich}}\footnote{\small {Also at
 the Institute for Advanced Study, Princeton NJ 08540 USA}}\\
 {\small{Geological and Planetary Sciences Division, CalTech
 Pasadena CA 91125 USA}}\\ {\small{e-mail: pmg @ sns.ias.edu~}} } \maketitle
 \begin{abstract}
In most books the Delaunay and Lagrange equations for the orbital
elements are derived by the Hamilton-Jacobi method: one begins
with the two-body Hamilton equations in spherical coordinates,
performs a canonical transformation to the orbital elements, and
obtains the Delaunay system. A standard trick is then used to
generalise the approach to the N-body case. We re-examine this
step and demonstrate that it contains an implicit condition which
restricts the dynamics to a  9(N-1)-dimensional submanifold of the
12(N-1)-dimensional space spanned by the elements and their time
derivatives. The tacit condition is equivalent to the constraint
that Lagrange imposed "by hand" to remove the excessive freedom,
when he was deriving his system of equations by variation of
parameters. It is the condition of the orbital elements being
osculating, i.e., of the instantaneous ellipse (or hyperbola)
being always tangential to the physical velocity. Imposure of any
supplementary condition different from the Lagrange constraint
(but compatible with the equations of motion) is legitimate and
will not alter the physical trajectory or velocity (though will
alter the mathematical form of the planetary equations).

This freedom of nomination of the supplementary constraint reveals
a gauge-type internal symmetry instilled into the equations of
celestial mechanics. Existence of this internal symmetry has
consequences for the stability of numerical integrators. Another
important aspect of this freedom is that any gauge different from
that of Lagrange makes the Delaunay system non-canonical. In a
more general setting, when the disturbance depends not only upon
positions but also upon velocities, there is a "generalised Lagrange
gauge" wherein the Delaunay system is symplectic. This special gauge
renders orbital elements that are osculating in the phase space. It
coincides with the regular Lagrange gauge when the perturbation is
velocity-independent.
% The latter case, however, cannot be described by the
% Hamilton-Jacobi method, because this method has the Lagrange
% constraint stiffly embedded into it.
\\
\\
% KEY WORDS: ~~~~Orbit integration, Lagrange system, Delaunay system,\\
% {${\left. \; \; \right.}^{\left. \; \right.}\;$}~~~~~~~~~~~~~~~~~~~~~Hidden symmetries, Gauge invariance.\\
% RUNNING HEAD: ~Hidden symmetry of the Lagrange and Delaunay systems
\end{abstract}

% \normalfont

% \pagebreak

\section{Euler and Lagrange}

\subsection{The history}

The planetary equations, which describe the evolution of the
orbital elements, constitute the cornerstone of the celestial
mechanics. Description of orbits in the language of Keplerian
elements (rather than in terms of the Cartesian coordinates) is
not only physically illustrative but also provides the sole means
for analysis of resonant interactions. These equations exist in a
variety of equivalent forms (those of Lagrange, Delaunay, Gauss,
Poincare) and can be derived by several different methods.

The earliest sketch of the method dates back to Euler's paper of
1748, which addresses the perturbations exerted upon one another
by Saturn and Jupiter. In the publication on the Lunar motion,
dated by 1753, Euler derived the equations for the longitude of
the node, $\;\Omega\;$, the inclination, $\;i\;$, and the quantity
$\;p\,\equiv\,a\,(1\,-\,e^2)\;$. Time derivatives of these three
elements were expressed through the components of the disturbing
force. Sixty years later the method was amended by Gauss who wrote
down similar equations for the other three elements and, thus,
obtained what we now call the Gauss system of planetary equations.
(The history of this scientific endeavour was studied by Subbotin
(1958), who insists that the Gauss system of planetary equations
should rather be called Euler system.). A modern but still
elementary derivation of this system belongs to Burns (1976).

In his m{\'e}moires of 1778, which received an honourable prize
from the Acad{\'e}mie des Sciences of Paris, Lagrange employed the
method of variation of parameters (VOP) to express the time
derivatives of the orbital elements through the disturbing
functions' partial derivatives with respect to the Cartesian
coordinates. In his m{\'e}moire of 1783 Lagrange furthered this
approach, while in Lagrange (1808, 1809, 1810) these equations
acquired their final, closed, shape: they expressed the orbital
elements' evolution in terms of the disturbing potentials'
derivatives with respect to the elements. Lagrange's derivation
rested upon an explicit imposure of the osculation condition,
i.e., of a supplementary vector equation (called the Lagrange
constraint) which guaranteed that the instantaneous ellipses (in
the case of bound motions) or hyperbolae (in the case of flyby
ones) are always tangential to the physical trajectory. Though it
has been long known (and, very possibly, appreciated by Lagrange
himself) that the choice of the supplementary conditions is
essentially arbitrary, and though a couple of particular examples
of use of non-osculating elements appeared in the literature
(Goldreich 1965; Brumberg et al 1971; Borderies \& Longaretti
1987), a comprehensive study of the associated freedom has not
appeared until very recently (Efroimsky 2002, 2003).

In the middle of the 19{\it{th}} century Jacobi applied a
canonical-transformation-based procedure (presently known as the
Hamilton-Jacobi approach) to the orbital dynamics, and offered a
method of deriving the Lagrange system. This technique is
currently regarded standard and is offered in many books. Though
the mathematical correctness of the Hamilton-Jacobi method is
beyond doubt, its application to celestial mechanics contains an
aspect that has long been overlooked (at least, in the
astronomical literature). This overlooked question is as follows:
where in the Hamilton-Jacobi derivation of the planetary equations
is the Lagrange constraint tacitly imposed, and what happens if we
impose a different constraint? This issue will be addressed in our
paper.

\subsection{The gauge freedom}

Mathematically, we shall concentrate on the N-body problem of
celestial mechanics, problem that for each body can be set as
 \be {\bf \ddot{\vec
r}}\;+\;\frac{\mu}{r^2}\;\frac{{\bf\vec r }}{r}\;=\;{\Delta
\bf{\vec F}}\;\;\;,
 \label{1}
 \ee
${\Delta \bf{\vec F}}\;$ being the disturbing force that vanishes
in the (reduced) two-body case, $\;\erbold\;$ being the position
relative to the primary, and $\;\mu\;$ standing for
$\;G(m_{planet}\,+\,m_{sun}\;$. A solution to the unperturbed
problem is a Keplerian ellipse (or hyperbola)
 \be
 {\bf \vec r }\;=\;{\bf \vec f} \left(C_1, ... , C_6, \,t \right)\;
 \label{2}
 \ee
parametrised by six constants (which may be, for example, the
Kepler or Delaunay elements). In the framework of the VOP approach
it gives birth to the ansatz
 \be
{\bf \vec r }\;=\;{\bf \vec f} \left(C_1(t), ... , C_6(t), \,t
\right)\,\;\;\;,
 \label{3}
 \ee
 the "constants" now being time-dependent and the functional form of $\;\bf{\vec{f}}\;$
 remaining the same as in (\ref{2}). Substitution of (\ref{3})
 into (\ref{1}) results in three scalar equations for six
 independent functions $\;C_i(t)\;$. In order to make the problem
 defined, Lagrange applied three extra conditions
   \be
\sum_i \;\frac{\partial {\bf \vec f}}{\partial C_i}\;\frac{d
C_i}{d t}\;=\; 0\;\;\;,\;
 \label{4}
  \ee
 that are often referred to as "the Lagrange constraint."  This
 constraint guarantees osculation, i.e.,
 that the functional dependence of the perturbed velocity upon the "constants"
 is the same as that of the unperturbed one. This happens because the physical
 velocity is
 \be
 {\bf{\dot{\erbold}}}\;=\;{\bf\vec g} \;+\;\sum_i \;\frac{\partial {\bf \vec f}}{\partial C_i}\;
 \frac{d C_i}{d t}\;
 \label{10}
 \ee
where $\;\bf\vec g\;$ stands for the unperturbed velocity that
emerged in the two-body setting. This velocity is, by definition,
a partial derivative of $\;\bf\vec f\;$ with respect to the last
variable:
 \be
 {\bf\vec g}\left(C_1,\,...\,,\,C_6,\,t \right)
 \;\equiv\;\frac{\partial {\bf\vec f}\left(C_1,\,...\,,\,C_6,\,t \right)}{\partial
 t}\;\;\;.
 \label{unperturbed_velocity}
 \ee
 This choice of
 supplementary conditions is convenient, but not at all necessary.
 A choice of any other three scalar relations (consistent with one
 another and with the equations of motion) will give the same
 physical trajectory, even though the appropriate solution for
 non-osculating variables $\;C_i\;$ will differ from the solution
 for osculating ones.

Efroimsky (2002, 2003) suggested to relax the Lagrange condition
and to consider
  \be
\sum_i \;\frac{\partial {\bf \vec f}}{\partial C_i}\;\frac{d
C_i}{d t}\;=\; {\bf {\vec
\Phi}}(C_{1}\,,\,...\,,\,C_6\,,\,t)\;\;\;,\;
 \label{13}
  \ee
 $\Phibold\,$  being an arbitrary function of time,
"constants" $\;C_i\;$ and their time derivatives of all orders.
For practical reasons it is convenient to restrict $\Phibold\,$ to
a class of functions that depend upon the time and the "constants"
only. (The dependence upon derivatives would yield
higher-than-first-order time derivatives of the $\;C_i\;$ in the
subsequent developments, which would require additional initial
conditions, beyond those on ${\bf\vec r}$ and ${\bf\dot{\vec
r}}\,$, to be specified in order to close the system.) Different
choices of $\Phibold\,$ entail different forms of equations for
$\;C_i\;$ and, therefore, different mathematical solutions in
terms of these "constants." A transition from one such solution to
another will, though, be a mere reparametrisation of the orbit.
The physical orbit itself will remain invariant. Such invariance
of the physical content of a theory under its mathematical
reparametrisations is called gauge symmetry. On the one hand, it
is in a close analogy with the gradient invariance of the Maxwell
electrodynamics and other field theories. On the other hand, it
illustrates some general mathematical structure emerging in the
ODE theory (Newman \& Efroimsky 2003).

If the Lagrange gauge (\ref{4}) is fixed, the parameters obey the
equation
 \be
\sum_j\;[C_n\;C_j]\;\frac{dC_j}{dt}\;=\;
 \frac{\partial \bf {\vec f}}{\partial C_n}\;
  {\Delta \bf{\vec F}}\;\;\;\;\;,
 \label{14}
 \ee
$[C_n\;C_j]\;$ standing for the unperturbed (i.e., defined as in
the two-body case) Lagrange brackets:
 \be
[C_n\;C_j]\;\equiv\;\frac{\partial {{\bf{\vec f}}}}{\partial
C_n}\, \frac{\partial {\bf {{\vec g}}}}{\partial
C_j}\,-\,\frac{\partial {{\bf{\vec f}}}}{\partial C_j}\,
\frac{\partial {\bf {{\vec g}}}}{\partial C_n}\;\;\;\;.
 \label{15}
 \ee
  To arrive at formula (\ref{14}), one should, according to Lagrange
   (1778, 1783, 1808, 1809), differentiate (\ref{10}), insert the outcome
   into (\ref{1}), and then combine the result with the Lagrange
   constraint (\ref{4}). (In the modern literature, this derivation
   can be found, for example, in Brouwer and Clemence (1961),
   Efroimsky (2002, 2003), Newman and Efroimsky (2003), Efroimsky
   and Goldreich (2003).)

In the simplest case the perturbing force depends only upon
positions and is conservative: $\;{\Delta \bf{\vec
F}}\;=\;{\partial R(\erbold)}/{\partial \erbold}\;$. Then the
right-hand side of (\ref{14}) will reduce to the partial
derivative of the disturbing function $\;R(\erbold )\;$ with
respect to $\;C_n\;$, whereafter inversion of the Lagrange-bracket
matrix will entail the Lagrange system of planetary equations (for
$\;C_i\;$ being the Kepler elements) or the Delaunay system (for
the parameters chosen as the Delaunay elements).

As explained in Efroimsky (2003), in an arbitrary gauge $\Phibold$
 equation (\ref{14}) will generalise to its gauge-invariant form
 \be
\sum_j\;[C_n\;C_j]\;\frac{dC_j}{dt}\;=\;
 \frac{\partial \bf {\vec f}}{\partial C_n}\;
  {\Delta \bf{\vec F}}\;-\; \frac{\partial{\bf
{\vec f}}}{\partial C_n} \;\frac{d \bf {\vec \Phi}}{dt}\;-\;
\frac{\partial  \bf {\vec g}}{\partial C_n} \;{\bf {\vec \Phi}}
\;\;\;\;,
 \label{16}
 \label{217}
 \ee
the Lagrange brackets $\;[C_n\;C_j]\;$ being still defined through
(\ref{15}). If we agree that $\;\Phibold\;$ is a function of both
time and the parameters $\;C_i\;$, but not of their derivatives,
then the right-hand side of (\ref{217}) will implicitly contain
the first time derivatives of $\;C_i\;$. It will then be
reasonable to move them into the left-hand side. Hence,
(\ref{217}) will be reshaped into
 \be
\sum_j\;\left(\,[C_n\;C_j]\;+\;\frac{\partial \bf\vec f}{\partial
C_n}\;\frac{\partial \bf {\vec \Phi}}{\partial C_j}\;
\,\right)\,\frac{dC_j}{dt}\;=\; \frac{\partial \bf {\vec
f}}{\partial C_n}\; {\Delta \bf{\vec F}}\;-\; \frac{\partial{\bf
{\vec f}}}{\partial C_n} \;\frac{\partial \bf {\vec
\Phi}}{\partial t}\;-\; \frac{\partial \bf {\vec g}}{\partial C_n}
\;{\bf {\vec \Phi}} \;\;\;\;.
 \label{general_F}
 \label{17}
 \ee
This is the general form of the gauge-invariant perturbation
equations of celestial mechanics, that follows from the VOP
method, for an arbitrary disturbing force $\;{\Delta \bf{\vec
F}}(\erbold,\,{\bf{\dot{\erbold}}},\,t)\,$ and under the
simplifying assumption that the arbitrary gauge function
$\;\Phibold\;$ is chosen to depend on the time and the parameters
$\;C_i\;$, but not on their derivatives.

For performing further algebraic developments of (\ref{16}) and
(\ref{17}), let us decide what sort of interactions will fall
within the realm of our interest. On general grounds, it would be
desirable to deal with forces that permit description in the
language of Lagrangians and Hamiltonians.

% \pagebreak

\section{Delaunay}

\subsection{Perturbations of Lagrangians and Hamiltonians}

Contributions to the disturbing force $\,\Delta \bf \vec F\,$
generally consist of two types, physical and inertial.  Inputs can
depend upon velocity as well as upon positions. As motivation for
this generalization we consider two practical examples. One is the
problem of orbital motion around a precessing planet: the orbital
elements are defined in the precessing frame, while the
velocity-dependent fictitious forces play the role of the
perturbation (Goldreich 1965, Brumberg et al 1971, Efroimsky \&
Goldreich 2003). Another example is the relativistic two-body
problem where the relativistic correction to the force is a
function of both velocity and position, as explained, for example,
in Brumberg (1992) and Klioner \& Kopeikin (1994). (It turns out
that in relativistic dynamics even the two-body problem is
disturbed, the relativistic correction acting as disturbance. This
yields the gauge symmetry that will cause ambiguity in defining
the orbital elements of a binary.) Finally, we shall permit the
disturbances to bear an explicit time dependence. Such a level of
generality will enable us to employ our formalism in noninertial
coordinate systems.

Let the unperturbed Lagrangian be $\;{\bf \dot{\vec
r}}^{\left.\,\right. 2}/2\,-\,U({\bf \vec r})$.  The disturbed
motion will be described by the new, perturbed, Lagrangian
 \ba
 {\cal L}\;=\;\frac{{\bf{\dot {\vec r}}}^{\left.\,\right. 2}}{2}\;-\;
  U({\bf \vec r})\;+
  \;\Delta {\cal L} ( {\bf \vec r},
 \,{ \bf { \dot { \vec {r}}}} ,\,t) \;\;\;,
 ~~~~~~~~~~~~~~~~~~~~~~~~~~~~~~~~~~~~~~~~
 \label{101}
 \ea
and the appropriately perturbed canonical momentum and
Hamiltonian:
 \ba
 {\bf {\vec p}}\;=\;{\bf{\dot {\vec r}}}\;+\;\frac{\partial \,\Delta {\cal L}}
{\partial {\bf{\dot
 {\vec r}}}} \;\;\;\;,
 ~~~~~~~~~~~~~~~~~~~~~~~~~~~~~~~~~~~~~~~~~~~~~~~~~~~~~~~~~~~
 \label{102}\\
 \nonumber\\
 {\cal H}\;=\;{\bf \vec p}\;{\bf {\dot {\vec r}}}\;-\;{\cal L}\;=\;\frac{{\bf \vec p}^{\left.
\,\right. 2}}{2}\;+ \;U\;+\;\Delta
 {\cal H}\;\;\;,\;\;\;\;\;\;\;\Delta {\cal H}\;\equiv\;-\;\Delta
 {\cal L}\;-\;\frac{1}{2}\,\left(\frac{\partial \,\Delta {\cal L}}{\partial {\bf{\dot
{\vec r}}}}
 \right)^2\;\;.
 \label{103}
 \ea
The Euler-Lagrange equations written for the perturbed Lagrangian
(\ref{101}) are:
 \be
 {\bf{\ddot {\vec r}}}\;=\;-\;\frac{\partial U}{\partial {\bf{\vec r}}} \;+\;
\Delta
 {\bf \vec F}\;\;\;\;,
 \label{104}
 \ee
where the disturbing force is given by
 \be
 \Delta {\bf \vec F}\;\equiv\;\frac{\partial \,\Delta {\cal L}}{\partial
 {\bf \vec r}}\;-\;\frac{d}{dt}\,\left(\frac{\partial \,\Delta {\cal L}}{\partial
{\bf{\dot
 {\vec r}}}}\right)\;\;\;\;.
 \label{105}
 \ee
We see that in the absence of velocity dependence  the
perturbation of the Lagrangian plays the role of disturbing
function. Generally, though, the disturbing force is not equal to
the gradient of $\,\Delta {\cal L}\,$, but has an extra term
called into being by the  velocity dependence.

As we already mentioned, this setup is sufficiently generic. For
example, it is convenient for description of a satellite orbiting
a wobbling planet: the inertial forces, that emerge in the
planet-related noninertial frame, will nicely fit in the above
formalism.

It is worth emphasising once again that, in the case of
velocity-dependent disturbances, the disturbing force is equal
neither to the gradient of the Lagrangian's perturbation nor to
the gradient of negative Hamiltonian's perturbation. This is an
important thing to remember when comparing results obtained by
different techniques. For example, in Goldreich (1965) the word
"disturbing function" was used for the negative perturbation of
the Hamiltonian. For this reason, the gradient of so defined
disturbing function was not equal to the disturbing force. A
comprehensive comparison of the currently developed theory with
that offered in Goldreich (1965) will be presented in a separate
publication (Efroimsky \& Goldreich 2003), where we shall
demonstrate that the method used there was equivalent to fixing a
special gauge (one described below in subsection 2.3 of this
paper).

\subsection{Gauge-invariant planetary equations}

Insertion of the generic force (\ref{105}) into (\ref{217}) will
bring us:
  \be
\sum_j\;[C_n\;C_j]\;\frac{dC_j}{dt}\;=\;\frac{\partial \bf {\vec
f}}{\partial C_n}\; \frac{\partial \, \Delta \cal L}{\partial
\erbold}\;-\; \frac{\partial{\bf {\vec f}}}{\partial C_n}
\;\frac{d}{dt}\left({\bf{\vec \Phi}} \,+\,\frac{\partial\, \Delta
\cal
 L}{\partial \bf{\dot{\vec r}}}\right)\;-\; \frac{\partial \bf
{\vec g}}{\partial C_n} \;{\bf {\vec \Phi}} \;\;\;\;.
 \label{219}
 \ee
 If we recall that, for a velocity-dependent disturbance,
 \be
\frac{\partial \,\Delta \cal L}{\partial C_n} \;=\;\frac{\partial
\, \Delta \cal L }{\partial {{{\erbold}}}} \,\frac{\partial{{\bf
{\vec f}}}}{\partial C_n} \;+\;\frac{\partial \,\Delta \cal
L}{\partial {\bf {\dot {\erbold}}}^{
 \left.~\right.} } \,\frac{\partial \bf\dot{\vec r} }{\partial
 C_n}\;=\;
\frac{\partial \,\Delta \cal L}{\partial {{{\erbold}}}}
\,\frac{\partial{{\bf {\vec f}}}}{\partial C_n}
\;+\;\frac{\partial \Delta \cal L}{\partial {\bf {\dot
{\erbold}}}^{
 \left.~\right.} } \,\frac{\partial ({\bf{\vec g}}\,+\,\Phibold) }{\partial
 C_n}
\;\;\;,
 \label{220}
 \ee
 then equality (\ref{219}) will look like this:
 \ba
% \nonumber
\sum_j\;[C_n\;C_j]\;\frac{dC_j}{dt}\;=\;\frac{\partial \,\Delta
\cal L }{\partial C_n} \;-\;\frac{\partial \,\Delta \cal
L}{\partial {\bf {\dot {\erbold}}}^{
 \left.~\right.} } \,\;\frac{\partial \Phibold  }{\partial C_n}
\;-\; \frac{\partial{\bf{\vec f}}}{\partial C_n}
 \;\frac{d}{dt} \left(
  {\bf
{\vec \Phi}}\,+\,\frac{\partial \,\Delta \cal L}{\partial
\bf\dot{\vec r }}
 \right)\;-\; \frac{\partial
\bf {\vec g} }{ \partial C_n} \;\left({\bf {\vec \Phi } }\;+\;
\frac{\partial \,\Delta \cal L}{\partial \bf\dot{\vec r}}
\right)\;\;.\; \label{221}
 \ea
 After subsequent addition and subtraction of
 $\;
 (1/2) \partial (
 (
 \partial
 ( \Delta {\cal L} ) / \partial {\bf{\dot{\vec{r}}}}
 )^2
 )/ \partial C_n
 \;$ in the right-hand side,
 the gauge function $\;\Phibold\;$ will everywhere appear in the
 company of $\;+\;\partial (\Delta \cal L)/\partial {\bf\dot{\vec r}}
 \;$:
 \ba
 \nonumber
\sum_j\;\left(\;[C_n\;C_j]\;+
 \;\frac{\partial {\bf\vec f}}{\partial C_n}\;
  \frac{\partial }{\partial C_j}\;
  \left(\frac{\partial \,\Delta \cal
 L}{\partial {\bf\dot{\vec r}}}\;+\;{\Phibold}
  \right)\;\right)
  \frac{dC_j }{dt }\;\;=
~~~~~~~~~~~~~~~~~~~~~~~~~~~~~~~~~~~~~~~~~\\
 \label{2211}\\
 \nonumber
\frac{\partial }{\partial C_n}\,\left[\Delta {\cal
L}\,+\,\frac{1}{2}\,\left(\frac{\partial \,\Delta \cal L}{\partial
\bf \dot{\vec r}} \right)^2 \right]
 \;-\;
\left( \frac{\partial \bf \vec g}{\partial C_n}\;+\;\frac{\partial
\bf \vec f}{\partial C_n}\;\frac{\partial}{\partial
t}\;+\;\frac{\partial \,\Delta \cal L}{\partial \bf\dot{\vec
r}}\;\frac{\partial }{\partial C_n} \right)\left(
  {\bf
{\vec \Phi}}\,+\,\frac{\partial \,\Delta \cal L}{\partial
\bf\dot{\vec r }}
 \right)
 \;\;\;,\;\;
 \ea
the sum in square brackets being equal to $\;-\,\Delta \cal H$.
While (\ref{general_F}) expressed the VOP method in the most
generic form it can have in terms of disturbing forces $\Delta
{\bf\vec F}({\bf\vec r},\,{\bf\dot{\vec r}},\,t)$, equation
(\ref{2211}) furnishes the most general form in terms of the
Lagrangian perturbation $\Delta {\cal L}({\bf\vec
r},\,{\bf\dot{\vec r}},\,t)$ (under the simplifying assumption
that the arbitrary gauge function $\;\Phibold\;$ is set to depend
only upon the time and the parameters $\;C_i\;$, but not upon
their derivatives).

The Lagrange brackets in (\ref{221}) are gauge-invariant; they
contain only functions $\,\bf\vec f\,$ and $\,\bf\vec g\,$ that
were defined in the unperturbed, two-body, setting. This enables
us to exploit the well-known expressions for the inverse of this
matrix. These look most simple (and are either zero or unity) in
the case when one chooses as the "constants" the Delaunay set of
orbital variables. As well known, this simplicity of the Lagrange
and their inverse, Poisson, brackets of the Delaunay elements is
the proof of these elements' canonicity in the unperturbed,
two-body, problem. When only a position-dependent disturbing
function $\;R({\bf\vec r})\,=\,\Delta {\cal L}({\bf\vec r})\;$ is
"turned on," the Delaunay elements still remain canonical,
provided the Lagrange gauge is imposed. This happens because, as
is well known (Brouwer \& Clemence 1961), the equations of motion
together with the Lagrange constraint yield, in that case, the
following equation:
 \ba
\sum_j\;[C_n\;C_j]\;\frac{dC_j}{dt}\;=\;\;\frac{\partial\, \Delta
\cal L }{\partial C_n}\,\;\;,\;\;\;\; \Delta {\cal L}\;=\;\Delta
{\cal L}\left({\bf\vec f}(C_1,\,...\,,\,C_6\,,\,t)\,\right)\;=\;
R\left({\bf\vec f}(C_1,\,...\,,\,C_6\,,\,t)\,\right)\;,
 \label{simple_case}
 \ea
which, in its turn, results in the standard Delaunay system.

In our case, though, the perturbation depends also upon
velocities; beside this, the gauge $\;\Phibold\;$ is set
arbitrary. Then our equation (\ref{2211}) will entail the
gauge-invariant Lagrange-type and Delaunay-type systems of
equations that are presented in Appendix 1. Interestingly, the
gauge-invariant Delaunay-type system is, generally,
non-symplectic. It regains the canonical form only in one special
gauge considered below (a gauge which coincides with the Lagrange
gauge when the perturbation bears no velocity dependence). This
can be proven either by a direct substitution of that special
gauge condition into the gauge-invariant Delaunay-type system
given in Appendix 1. An easier way would be to fix the gauge
already in (\ref{2211}), and this is what we shall do in the next
subsection.

\subsection{The generalised Lagrange gauge:\\
 gauge wherein the Delaunay-type system becomes canonical}

We transformed (\ref{219}) into (\ref{2211}) for two reasons: to
single out the negative perturbation of the Hamiltonian and to
reveal the advantages of the gauge
 \ba
 \Phibold\;=\;-\;\frac{\partial\,\Delta L}{\partial
 {\bf\dot{\vec r}}}\;\;\;,
 \label{special_gauge}
 \ea
which reduces to $\Phibold=0$ for velocity-independent
perturbations. The first remarkable peculiarity of
(\ref{special_gauge}) is that in this gauge the canonical momentum
$\,\bf \vec p\,$ is equal to $\,\bf \vec g\,$ (as can be seen from
(\ref{10}) and (\ref{102})):
 \be
 {\bf\vec g}\;=\;{\bf\dot{\vec r}}\;-\;\Phibold\;=\;{\bf\dot{\vec
  r}}\;+\;\frac{\partial \Delta \cal L}{\partial \bf\dot{\vec r}}
  \;=\;{\bf\vec p}\;\;\;.
 \label{gp}
 \ee
We see that in this gauge not the velocity but the momentum in the
disturbed setting is the same function of time and "constants" as
it used to be in the unperturbed, two-body, case. Stated
differently, the instantaneous ellipses (hyperbolae) defined in
this gauge will osculate the orbit {\bf in the phase space}. For
this reason our special gauge (\ref{special_gauge}) will be called
the "generalised Lagrange gauge."

Another good feature of (\ref{special_gauge}) is that in this
gauge equation (\ref{2211}) acquires an especially simple form:
 \ba
\sum_j\;[C_n\;C_j]\;\frac{dC_j}{dt}\;=\;-\;\frac{\partial\;\Delta
\cal H }{\partial C_n}\,\;\;
 \label{simple_form}
 \ea
whose advantage lies not only in its brevity, but also in the
invertibility of the matrix emerging in its left-hand side. As
already mentioned above, the gauge invariance of definition
(\ref{15}) enables us to employ the standard (Lagrange-gauge)
expressions for $[C_n\,C_j]^{-1}$ and, thus, to get the planetary
equations by inverting matrix $[C_n\,C_j]$ in (\ref{simple_form}).
The resulting gauge-invariant Lagrange- and Delaunay-type systems
are presented in Appendix 1.

In the special gauge (\ref{special_gauge}), however, the situation
is much better. Comparing (\ref{simple_case}) with
(\ref{simple_form}), we see that in the general case of an
arbitrary $R=\Delta {\cal L}({\bf\vec r},\,{\bf\dot{\vec r}},\,t)$
one arrives from (\ref{simple_form}) to the same equations as from
(\ref{simple_case}), except that now they will contain
$\,-\,\Delta \cal H$ instead of $R=\Delta \cal L$. These will be
the Delaunay-type equation in the generalized Lagrange gauge:
 \ba
 \frac{dL}{dt}=
 \frac{\partial \,\Delta {\cal H}}{\partial (\,-\,M_o)}
  \;\;,\;\;\;\;\;\;
 \frac{d(\,-\,M_o)}{dt}\;=\,-\,\frac{\partial  \,\Delta {\cal H}}{\partial
L}\;\;\;\;,\;\;\;\;\;\;\;
 \label{Delaunay.12}
 \ea
 \ba
 \frac{dG}{dt}\;=\;\frac{\partial  \,\Delta {\cal H}}{\partial
(\,-\,\omega )}\;\;\;\;,\;\;\;\;\;\; \frac{d(\,-\,\omega)
}{dt}\;=\,-\,\frac{\partial  \,\Delta {\cal H}}{\partial
G}\,\;\;\;\;,\;\;\;\;\;
 \label{Delaunay.34}
 \ea
 \ba
 \frac{d H}{dt}\,=\,\frac{\partial  \,\Delta {\cal H}}{\partial
(\,-\,\Omega )}\,\;\;\;\;,\;\;\;\;\;\; \frac{d (\,-\,\Omega
)}{dt}\,=\,-\,\frac{\partial
 \,\Delta {\cal H}}{\partial H}\,\;\;\;\;.\;\;\;\;\;\;\;
 \label{Delaunay.56}
 \ea
where
 \ba L\,\equiv\,\mu^{1/2}\,a^{1/2}\,\;\;,\;\;\;\;\;
G\,\equiv\,\mu^{1/2}\,a^{1/2}\,\left(1\,-\,e^2\right)^{1/2}\,\;\;,\;\;\;\;
H\,\equiv\,\mu^{1/2}\,a^{1/2}\,\left(1\,-\,e^2\right)^{1/2}\,\cos
\inc\,\;\;\;\;\;.\;\;
 \label{229}
 \ea
We see that in this special gauge the
Delaunay-type equations indeed become canonical, and the role of
the effective new Hamiltonian is played exactly by the Hamiltonian
perturbation which emerged earlier in (\ref{103}).

Thus we have proven an interesting {\textbf{THEOREM$\,$}}:
 {\textbf{~Even though the gauge-invariant Delaunay-type system
 (\ref{Delaunay.1} - \ref{Delaunay.6}) is not generally canonical, it
 becomes canonical in one special gauge (\ref{special_gauge}) which we
 call the "generalised Lagrange gauge."}} This theorem can be proved
 in a purely Hamiltonian language, as is done in Section 3.3.

\section{Hamilton and Jacobi}

\subsection{The concept}

A totally different approach to derivation of the planetary
equations is furnished by the technique worked out in 1834 - 1835
by Hamilton and refined several years later by Jacobi. In the
lecture course shaped by 1842 and published as a book in 1866,
Jacobi formulated his famous theorem and applied it to the
celestial motions. Jacobi chose orbital elements that were some
combinations of the Keplerian ones. His planetary equations can be
easily transformed into the Lagrange system by the differentiation
chain rule (Subbotin 1968). Later authors used this method for a
direct derivation of the Lagrange and Delaunay systems, and thus
the Hamilton-Jacobi approach became a part and parcel of almost
any course in celestial mechanics. To some of these sources we
shall refer below. The full list of pertinent references would be
endless, so it is easier to single out a couple of books that
break the code by offering alternative proofs: these exceptions
are Kaula (1968) and Brouwer and Clemence (1961).

Brouwer and Clemence (1961) use the VOP method (like in Lagrange
(1808, 1809, 1810)) which makes the imposition of the Lagrange
constraint explicit. Kaula (1968) undertakes, by means of the
differentiation chain rule, a direct transition from the Hamilton
equations in a Cartesian frame to those in terms of orbital
elements. As explained in Efroimsky (2002, 2003), in Kaula's
treatment the Lagrange constraint was imposed tacitly.

It is far less easy to understand where the implicit gauge fixing
is used in the Hamilton-Jacobi technique. This subtlety of the
Hamilton-Jacobi method is so well camouflaged that through the
century and a half of the method's life this detail has never been
brought to light (at least, in the astronomical literature). The
necessity of such a constraint is evident: one has to choose one
out of infinitely many sets of orbital elements describing the
physical trajectory. Typically, one prefers the set of orbital
elements that osculates with the trajectory, so that the physical
orbit be always tangential to the instantaneous ellipse, in the
case of bound orbits, or to the instantaneous hyperbola, in the
case of flybys. This point is most easily illustrated by the
following simple example depicted on Fig.1. Consider two coplanar
ellipses with one common focus. Let both ellipses rotate, in the
same direction within their plane, about the shared focus; and let
us assume that the rotation of one ellipse is faster than that of
the other. Now imagine that a planet is located at one of the
points of these ellipses' intersection, and that the rotation of
the ellipses is such that the planet is always at the
instantaneous point of their intersection. One observer will say
that the planet is swiftly moving along the slower rotating
ellipse, while another observer will argue that the planet is
slowly moving along the fast-rotating ellipse. Both viewpoints are
acceptable, because one can divide, in an infinite number of ways,
the actual motion of the planet into a motion along some ellipse
and a simultaneous evolution of that ellipse. The Lagrange
constraint (\ref{4}) singles out, of all the multitude of evolving
ellipses, that unique ellipse which is always tangential to the
total (physical) velocity of the body. This way of gauge fixing is
natural but not necessary. Besides, as we already mentioned, the
chosen gauge (\ref{4}) will not be preserved in the course of
numerical computations. (Sometimes osculating elements do not
render an intuitive picture of the motion. In such situations
other elements are preferred. One such example is a circular orbit
about an oblate planet. The osculating ellipse precesses with the
angular velocity of the satellite, and its eccentricity is
proportional to the oblateness factor $J_2$. Under these
circumstances the so-called geometric elements are more convenient
than the osculating ones (Borderies \& Longaretti 1987).)

We remind the reader that the Hamilton-Jacobi treatment is based
on the simple facts that the same motion can be described by
different mutually interconnected canonical sets
$\;(\,q,\,p,\,{\cal H}(q,p)\,)\;$ and $\;(\,Q,\,P,\,{\cal
H}^{*}(Q,P)\,)\;$, and that fulfilment of the Hamilton equations
along the trajectory makes the infinitesimally small quantities
 \be
d \theta \,=\,p\,dq\,-\,{\cal H}\,dt
 \label{25}
 \ee
and
 \be
d \tilde{\theta} \,=\,P\,dQ\,-\,{\cal H}^{*}\,dt
 \label{26}
 \ee
perfect differentials. Subtraction of the former from the latter
shows that their difference,
 \be
 -\,dW\,\equiv\,d \tilde{\theta}\,-\,d
 \theta\;=\;P\,dQ\,-\,p\,dq\,-\,\left({\cal H}^{*}\,-\,{\cal H}\right)\,dt
 \label{27}
 \ee
is a perfect differential, too. Here $\;q,\,p,\,Q,\,$ and $\,P\,$
contain N components each. If we start with a system described by
$\;(\,q,\,p,\,{\cal H}(q,p)\,)\;$, it is worth looking for such a
re-parametrisation $\;(\,Q,\,P,\,{\cal H}^{*}(Q,P)\,)\;$ that the
new Hamiltonian $\;H^{*}\;$ is constant in time, because in these
variables the canonical equations simplify. Especially convenient
is to find a transformation that nullifies the new Hamiltonian
$\;{\cal H}^{*}\;$, for in this case the new canonical equations
will render
 \begin{center}
        \epsfxsize=110mm
        \epsfbox{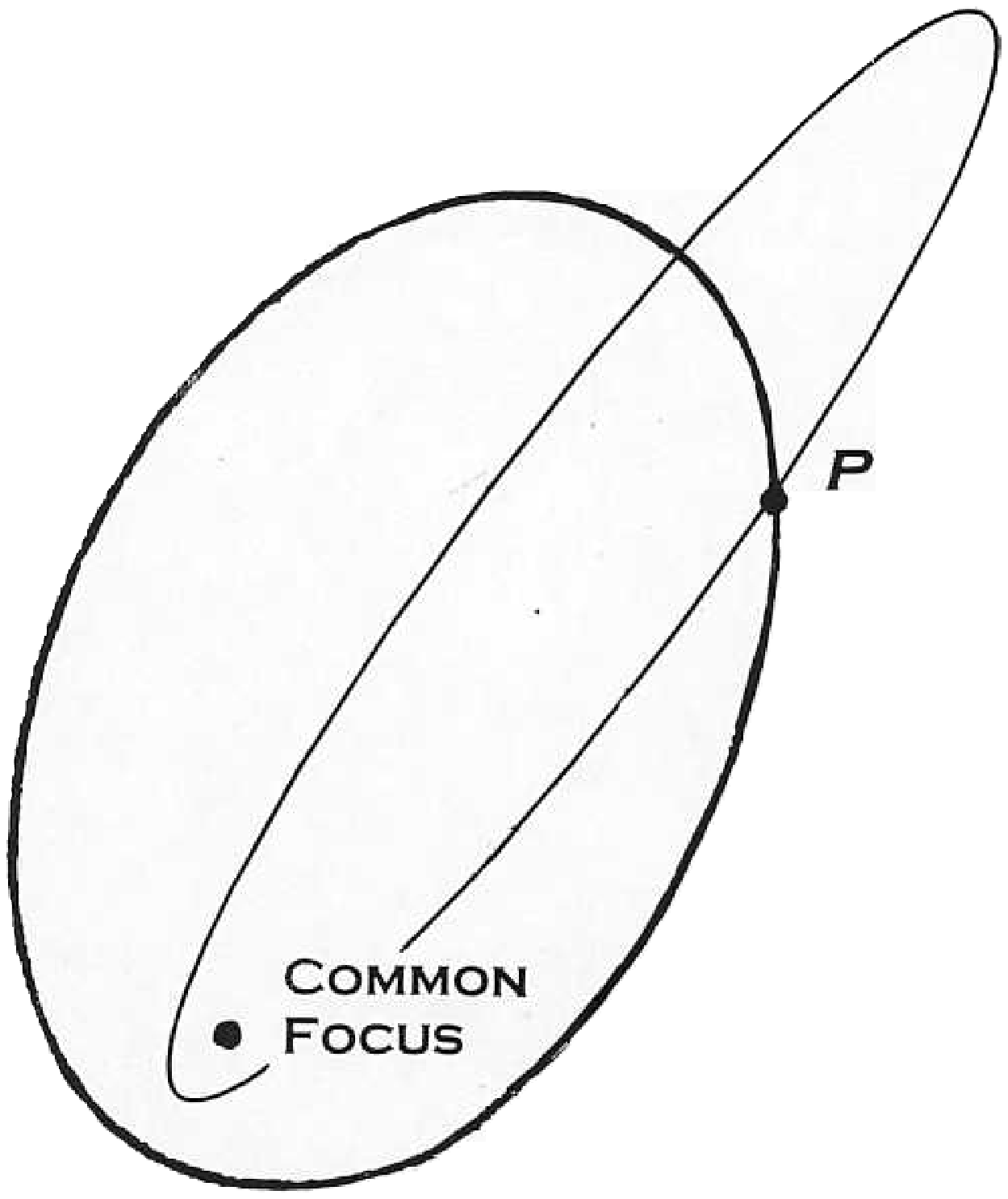}
  \end{center}
  \mbox{\small
 \parbox[b]{5.5in}{{\underline{Fig.1.}}~~{These two coplanar ellipses share one of their foci and are
assumed to rotate about this common focus in the same direction,
always remaining within their plane. Suppose that the rotation of
one ellipse is faster than that of the other, and that a planet is
located at one of the points of these ellipses' intersection, P,
and that the rotation of the ellipses is such that the planet is
always at the instantaneous point of their intersection. We may
say that the planet is swiftly moving along the slower rotating
ellipse, while it would be equally legitimate to state that the
planet is slowly moving along the fast-rotating ellipse. Both
interpretations are valid, because one can divide, in an infinite
number of ways, the actual motion of the planet into a motion
along some ellipse and a simultaneous evolution of that ellipse.
The Lagrange constraint (\ref{4}) singles out, of all the
multitude of evolving ellipses, that unique ellipse which is
always tangential to the total, physical, velocity of the planet.
 }}}\\

~\\

\noindent
the variables $\;(\,Q,\,P\,)\;$ constant. One way of
seeking such transformations is to consider $\;W\;$ as a function
of only $\;q\;$, $\;Q\;$, and $\;t\,$. Under this assertion, the
above equation will entail:
 \be
-\,\frac{\partial W}{\partial t}\;dt\;-\;\frac{\partial
W}{\partial Q}\;dQ\;-\;\frac{\partial W}{\partial q}\;dq\;=\;
P\,dQ\;-\;p\,dq\;+\;\left({\cal H}\,-\,{\cal H}^{*}\right)\,dt
 \label{28}
 \ee
whence
 \be
 P\;=\;-\;\frac{\partial W}{\partial
 Q}\;\;\;\;,\;\;\;\;\;\;\;p\;=\;\frac{\partial W}{\partial
 q}\;\;\;\;,\;\;\;\;\;\;\;{\cal H}\;+\;\frac{\partial W}{\partial
 t}\;=\;{\cal H}^{*}\;\;\;.
 \label{29}
 \ee
The function $\;W\;$ can be then found by solving the Jacobi
equation
 \be
{\cal H}\left(q,\,\frac{\partial W}{\partial
q}\,,\,t\right)\;+\;\frac{\partial W}{\partial t}\; =\;{\cal
H}^{*}
 \label{30}
 \ee
 where $\;{\cal H}^{*}\;$ is a constant. It is very convenient to make it
 equal to zero. Then the above equation can be easily solved in
 the unperturbed (reduced) two-body setting. This solution, which
 has long been known, is presented, in a very compact form,
 in Appendix 2. It turns out that, if the
 spherical coordinates and their conjugate momenta are taken as a
 starting point, then the eventual canonical variables
 $\;Q,\,P\;$ corresponding to $\;{\cal H}^{*}(Q,\,P)\,=\,0\;$ are the
 Delaunay elements:
 \ba
 \nonumber
 Q_1\;\equiv\;L\;=\;\sqrt{\mu\;a}\;\;\;\;\;\;\;\;\;\;\;\;\;\;\;\;\;\;\;\;\;\;\;\;\;,\;\;\;\;\;\;\;\;
 P_1\;=\;-\;M_o\;\;\;\;\;\;\;\;\;\;\;\;\;\;\;,
 \\
 \nonumber\\
 Q_2\;\equiv\;G\;=\;\sqrt{\mu\,a\,\left(1\,-\,e^2\right)}\;\;\;\;\;\;\;\;\;\;\;\;,\;\;\;\;\;\;\;
 P_2\;=\;-\;\omega\;\;\;\;\;\;\;\;\;\;\;\;\;\;\;\;\;\;,
 \label{45}\\
 \nonumber\\
 \nonumber
 Q_3\;\equiv\;H\;=\;\sqrt{\mu\,a\,\left(1\,-\,e^2\right)}\,\cos\,i\;\;\;\;,\;\;\;\;\;\;\;
 P_3\;=\;-\;\Omega\;\;\;\;\;\;\;\;\;\;\;\;\;\;\;\;\;\;.
 \ea

\subsection{Where can free cheese be found?}

The transition from two-body to N-body celestial mechanics
is presented in numerous books. However, none of them
explain how the Lagrange constraint is implicitly involved in the
formalism.

Before we move on, let us cast a look back at what has been
accomplished in the two-body case. We started out with a
Hamiltonian problem $\;(\,q,\,p,\,{\cal H}\,)\;$ and re-formulated
its equations of motion
 \be
 \dot q\;=\;\frac{\partial {\cal H}}{\partial p}\;\;\;,\;\;\;\;\;\;\dot
 p\;=\;-\;\frac{\partial {\cal H}}{\partial q}
 \label{46}
 \ee
in terms of another set $\;(\,Q,\,P,\,{\cal H}^{*}\,)\;$:
 \ba
 \nonumber
 q\;=\;\phi(Q,\,P,\,t)\\
 \label{47}\\
 \nonumber
 p\;=\;\psi(Q,\,P,\,t)
 \ea
 so that the above equations are mathematically equivalent to the
 new system
 \be
 \dot Q\;=\;\frac{\partial {\cal H}^{*}}{\partial P}\;\;\;,\;\;\;\;\;\;\dot
 P\;=\;-\;\frac{\partial {\cal H}^{*}}{\partial Q}\;\; .
 \label{48}
 \ee
The simple nature of the 2-body setting enabled us to carry out
this transition so that our new Hamiltonian $\;{\cal H}^{*}\;$
vanishes and the variables $\;Q\;$ and $\;P\;$ are, therefore,
constants. This was achieved by means of a
transformation-generating function $\;W(q,\,Q,\,t)\;$ obeying the
Jacobi equation (\ref{30}). After formula (\ref{42}) for this
function was written down, the explicit form of dependence
(\ref{47}) can be found through the relations $\;P\,=\,-\,\partial
W/\partial Q\;$. This is given by (\ref{45}).

To make this machinery function in an N-body setting, let us first
consider a disturbed two-body case. The number of degrees of
freedom is still the same (three coordinates $\,q\,$ and three
conjugate momenta $\,p\,$), but the initial Hamiltonian is
perturbed:
 \be
 \dot q\;=\;\frac{\partial ({\cal H}\,+\,\Delta {\cal H})}{\partial p}\;\;\;,\;\;\;\;\;\;\dot
 p\;=\;-\;\frac{\partial ({\cal H}\,+\,\Delta {\cal H})}{\partial q}\;\;.
 \label{49}
 \ee
Trying to implement the Hamilton-Jacobi method (\ref{28}) -
(\ref{30}), for the new Hamiltonians $\,({\cal H}\,+\,\Delta {\cal
H})\,$, $\;({\cal H}^{*}\,+\,\Delta {\cal H})\,$ and for some
generating function $\;V(q,\,Q,\,t)\;$, we shall arrive at
 \be
-\,\frac{\partial V}{\partial t}\;dt\;-\;\frac{\partial
V}{\partial Q}\;dQ\;-\;\frac{\partial V}{\partial q}\;dq\;=\;
P\,dQ\;-\;p\,dq\;+\;\left[\,({\cal H}\,+\,\Delta {\cal
H})\,-\,({\cal H}^{*}\,+\,\Delta {\cal
H})\,\right]\,dt\;\;\;,\;\;\;\;
 \label{50}
 \ee
 \be
 P\;=\;-\;\frac{\partial V}{\partial
 Q}\;\;\;\;,\;\;\;\;\;\;\;p\;=\;\frac{\partial V}{\partial
 q}\;\;\;\;,\;\;\;\;\;\;\;{\cal H}\;+\;\Delta {\cal H}\;+\;\frac{\partial V}{\partial
 t}\;=\;{\cal H}^{*}\;+\;\Delta {\cal H}\;\;,\;\;\;\;
 \label{51}
 \ee
 \be
 {\cal H}\left(q,\,\frac{\partial V}{\partial q}\,,\,t\right)\;
 +\;\frac{\partial V}{\partial t}\; =\;{\cal H}^{*}\;\;\;.\;\;\;\;
 \label{52}
 \ee
We see that $\,V\,$ obeys the same equation as $\,W\,$ and,
therefore, may be chosen to coincide with it. Hence, the new,
perturbed, solution $\;(q,\,p)\;$ will be expressed through the
perturbed "constants" $\;Q(t)\,$ and $\,P(t)\;$ in the same
fashion as the old, undisturbed, $\;q\;$ and $\;p\;$ were
expressed through the old constants $\;Q\;$ and $\;P\;$:
 \ba
 \nonumber
 q\;=\;\phi(Q(t),\,P(t),\,t)\;\;\;\;,\;\\
 \label{53}\\
 \nonumber
 p\;=\;\psi(Q(t),\,P(t),\,t)\;\;\;,\;
 \ea
$\phi\;$ and $\;\psi\;$ being the same functions as those in
(\ref{47}). Benefitting from this form-invariance, one can now
master the N-particle problem. To this end, one should choose the
transformation-generating function $\;V\;$ to be additive over the
particles, whereafter the content of subsection 3.1 shall be
repeated verbatim for each of the bodies, separately. In the end
of this endeavour one will arrive to $\,N-1\,$ Delaunay sets
similar to (\ref{45}), except that now these sets will be
constituted by {\bf{instantaneous}} orbital elements. The
extension of the preceding subsection's content to the N-body case
seems to be most straightforward and to involve no additional
assumptions. To dispel this illusion, two things should be
emphasised. One, self-evident, fact is that the quantities $\;Q\;$
and $\;P\;$ are no longer conserved after the disturbance
$\;\Delta {\cal H}\;$ is added to the zero Hamiltonian $\;{\cal
H}^{*}\;$. The second circumstance is that a change in a
Hamiltonian implies an appropriate alteration of the Lagrangian.
In the simple case of $\;\Delta {\cal H}\;$ being a function of
the coordinates and time only (not of velocities or momenta),
addition of $\;\Delta {\cal H}\;$ to the Hamiltonian implies
addition of its opposite to the Lagrangian. Since this extra term
bears no dependence upon velocities, the expressions for momenta
through the coordinates and time will stay form-invariant. Hence
(if the Lagrangian is not singular), the functional dependence of
the velocities upon the coordinates and momenta will, also,
preserve their functional form $\,v(q,\,p,\,t)\,$:
 \ba
 \nonumber
 without~~perturbation:~~~~~~~p\,\equiv\,\frac{\partial
{\cal L}(q,\dot{q},t)}{\partial \dot
q}\;\;\;\;\Longrightarrow\;\;\;~\dot{q}=v(q,p,t)\;\;\;,
~~~~~~~~~~~~~~~~~~~~~~~~~~~~~~~\\
 \label{54}\\
 \nonumber
with~~perturbation:~~~~p\equiv\frac{\partial\,\left(\, {\cal
L}(q,\,\dot{q},\,t)\,+\,\Delta {\cal L}(q,\,t)\,\right)}{\partial
\dot q}=\frac{\partial {\cal L}(q,\,\dot{q},\,t)}{\partial \dot
q}\;\;~~\Rightarrow~~\;\;~\dot{q}=v(q,\,p,\,t)\;\;\;,~~~~
 \ea
where the new, perturbed, dependence
$\;\dot{q}\;=\;v[\,q(\,Q(t),\,P(t),\,t),\,p(\,Q(t),\,P(t),\,t),\,t]\;$
has the same functional form as the old one,
$\;\dot{q}\;=\;v[\,q(\,Q,\,P,\,t),\,p(\,Q,\,P,\,t),\,t]\;$.
Together with (\ref{53}), this means that the dependence of the
new $\;\dot q\;$ upon the new $\;P(t)\;$ and $\;Q(t)\;$ will have
the same functional form as the dependence of the old $\;\dot q\;$
upon the constants $\;Q\;$ and $\;P\;$:
 \ba
 \frac{d}{dt}\;\;q(\,Q(t),\,P(t),\,t)\;=\;\frac{\partial}{\partial
 t} \;\;q(\,Q(t),\,P(t),\,t) \;\;\;.
 \label{55}
 \ea
In other words,
 \ba
 \sum_{i=1}^{6}\,\frac{\partial q}{\partial D_i}\;\dot
 D_i\;=\;0\;\;\;,
 \label{56}
 \ea
where $\;D_i\;$ denotes the set of perturbed variables
$\;(Q(t),\,P(t))\;$. In the astronomical applications, $\;D_i\;$
may stand for the Delaunay set.

This is the implicit condition under which the Hamilton-Jacobi
method works (in the above case of velocity-independent
disturbance). Violation of (\ref{56}) would invalidate our
cornerstone assumption (\ref{48}). This circumstance imposes a
severe restriction on the applicability of the Hamilton-Jacobi
theory. In the astronomical context, this means that the Delaunay
elements (\ref{45}) must be osculating. Indeed, if $\;D_i\;$
denote a set of orbital elements, then expression (\ref{56}) is
equivalent to the Lagrange constraint (\ref{4}) discussed in
Section 1. There the constraint was imposed upon the Keplerian
elements; however, its equivalence to (\ref{56}), which is written
in terms of the Delaunay variables, can be easily proven by the
differentiation chain rule.

\subsection{The case of momentum-dependent disturbances}

When the perturbation of the Lagrangian depends also upon
velocities (and, therefore, the Hamiltonian perturbation carries
dependence upon the canonical momenta), the special gauge
(\ref{special_gauge}) wherein the Delaunay-type system preserves
its canonicity differs from the Lagrange gauge. This was proven in
subsection 2.3 in the Lagrangian language. Now we shall study this
in Hamiltonian terms. Our explanation will be sufficiently general
and will surpass the celestial-mechanics setting. For this reason
we shall use notations $\;q,\,p\;$, not $\;{\bf\vec r},\,{\bf\vec
p}\;$. The development will, as ever, begin with an unperturbed
system described by canonical variables obeying
 \be
 \dot q\;=\;\frac{\partial {\cal H}}{\partial p}\;\;\;,\;\;\;\;\;\;\dot
 p\;=\;-\;\frac{\partial {\cal H}}{\partial q} \;\;\;\;.
 \label{A1}
 \ee
This dynamics may be re-formulated in terms of the new quantities
$\;(\,Q,\,P\,)\;$:
 \ba
 \nonumber
 q\;=\;\phi(Q,\,P,\,t)\\
 \label{A2}\\
 \nonumber
 p\;=\;\psi(Q,\,P,\,t)
 \ea
 so that the Hamiltonian equations (\ref{A1}) are equivalent to
 \be
 \dot Q\;=\;\frac{\partial {\cal H}^{*}}{\partial P}\;\;\;,\;\;\;\;\;\;\dot
 P\;=\;-\;\frac{\partial {\cal H}^{*}}{\partial Q}\;\;\;\;\;.
 \label{A3}
 \ee
For simplicity, we shall assume that $\;{\cal H}^{*}\;$ is zero.
Then the new canonical variables will play the role of adjustable
constants upon which the solution (\ref{A2}) of (\ref{A1})
depends.

We now wish to know under what circumstances a modified canonical
system
 \be
 \dot q\;=\;\frac{\partial ({\cal H}\,+\,\Delta {\cal H})}{\partial p}\;\;\;,\;\;\;\;\;\;\dot
 p\;=\;-\;\frac{\partial ({\cal H}\,+\,\Delta {\cal H})}{\partial
 q}\;\;\;\;,\;\;\;\;\;\;\;\Delta {\cal H}\;=\;\Delta {\cal H}(q,\,p,\,t)
 \label{A4}
 \ee
will be satisfied by the solution
 \ba
 \nonumber
 q\;=\;\phi(Q(t),\,P(t),\,t)\\
 \label{A5}\\
 \nonumber
 p\;=\;\psi(Q(t),\,P(t),\,t)
 \ea
of the same functional form as (\ref{A2}) but with time-dependent
parameters obeying
 \be
 \dot Q\;=\;\frac{\partial \,\Delta {\cal H}}{\partial P}\;\;\;,\;\;\;\;\;\;\dot
 P\;=\;-\;\frac{\partial \,\Delta {\cal H}}{\partial
 Q}\;\;\;\;.
 \label{A4}
 \ee
This situation is of a more general sort than that addressed in
subsection 3.2, in that the perturbation $\;\Delta {\cal H}\;$ now
depends also upon the momentum.

Under the assumption of (\ref{A2}) being (at least, locally)
invertible, substitution of the equalities
  \ba
  \dot{ Q}\;=\;\frac{\partial \,\Delta {\cal H} }{
  \partial P }\;=\;
  \frac{
  \partial \,\Delta {\cal H}}{\partial q }\;
  \frac{\partial  q }{
  \partial  P }
  \;+\;\frac{
  \partial \,\Delta {\cal H}}{\partial p }\;
  \frac{\partial p }{ \partial  P }
  \label{A5}
  \ea
 and
  \ba
  \dot{P }\;=\;-\;\frac{\partial \,\Delta {\cal H}}{\partial
  Q}\;=\;-\;\frac{\partial  \,\Delta {\cal H}
  }{
  \partial q }\;\frac{\partial q }{
  \partial Q  }\;-\;\frac{\partial  \,\Delta {\cal H} }{ \partial p
  } \;\frac{\partial p  }{ \partial Q
  }\;\;\;\;\;\;
  \label{A6}
  \ea
 into the expression for velocity
  \ba
  \dot{ q }\;=\;\frac{\partial q}{\partial t}\;+\;\frac{\partial q
  }{\partial Q }\;\dot{Q}\;+\;\frac{\partial q }{\partial P
  }\;\dot{P}\;
  \label{A7}
  \ea
 leads to
 \ba
 \dot{ q}\;=\;\frac{\partial q}{\partial t}\;+\;\left(
 \frac{\partial  q}{\partial  Q  }\;\frac{\partial  q}{\partial P
 }\;-\;\frac{\partial q }{\partial  P  }\;\frac{\partial q
 }{\partial Q }
 \right)\;\frac{\partial \,\Delta {\cal H}}{\partial q}\;+\; \left(
 \frac{\partial  q}{\partial  Q  }\;\frac{\partial  p}{\partial P
 }\;-\;\frac{\partial q }{\partial  P  }\;\frac{\partial p
 }{\partial Q }
 \right)\;\frac{\partial \,\Delta {\cal H}}{\partial p}\;\;.
 \label{A80}
 \ea
 Here the coefficient accompanying $\;{\partial \Delta \cal H}/{\partial q}\;$
 identically vanishes, while that accompanying $\;{\partial \Delta \cal H}/{\partial p}\;$
 coincides with the Jacobian of the canonical transformation and
 is, therefore, unity:
 \be
 \frac{\partial  q}{\partial  Q  }\;\frac{\partial  p}{\partial P
 }\;-\;\frac{\partial q }{\partial  P  }\;\frac{\partial p
 }{\partial Q }\;=\;1\;\;\;.
 \label{A9}
 \ee
 So if we introduce, in the spirit of (\ref{unperturbed_velocity}),
notation
 \be
 g\;\equiv\;\frac{\partial q}{\partial t}~~~,
 \label{A99}
 \ee
then (\ref{A80}) will lead to
 \ba
 \dot{q}\;=\;g\;+\;\left( \frac{\partial \, \Delta {\cal H} }{\partial p}
 \right)_{q,\,t}\  \;\;\;.\;
 \label{A8}
 \ea
Expression (\ref{A8}) establishes the link between the regular VOP
method and the canonical treatment. It shows that, to preserve the
symplectic description, one must always choose a particular gauge
$\;\Phi\;=\;{\partial \,\Delta {\cal H}}/{\partial p}\;$. Needless
to say, this is exactly the generalised Lagrange gauge
(\ref{special_gauge}) discussed in subsection 2.3.  A direct,
though very short, proof is as follows.

On the one hand, the Hamilton equation for the perturbed Hamiltonian
(\ref{103}) is:
 \ba
 \dot{q}\;=\;\frac{\partial \,\left({\cal H}\,+\,\Delta {\cal H}\right)}{\partial
 p}\;=\;p\;+\;\frac{\partial \,\Delta {\cal H}}{\partial
 p}\;
 \label{5558}
 \ea
while, on the other hand, the definition of momentum entails, for
the Lagrangian (\ref{101}):
 \ba
 p\,\equiv\,\frac{\partial\,\left(\,
{\cal L}(q,\,\dot{q},\,t)\,+\,\Delta {\cal
L}(q,\,\dot{q},\,t)\,\right)}{\partial \dot
q}\;=\;\dot{q}\;+\;\frac{\partial \,\Delta {\cal L}}{\partial
\dot{q}}\;\;.
 \label{5559}
 \ea
By comparing the latter with the former we arrive at:
 \ba
 \Phi\;\equiv\; \left( \frac{ \partial \, \Delta {\cal H} }{
\partial p}\right)_{q,\,t}\;=\;-\;\left(\frac{\partial\;\Delta \cal L}{\partial
 \dot{q}}\right)_{q,\,t}~~~~~~~~~~~~~~~~~~~~~~~~
 \label{5557}
 \ea
which coincides with (\ref{special_gauge}). Thus we see that
transformation (\ref{A2}) being canonical is equivalent to the
partition of the physical velocity $\;\dot q\;$ in a manner
prescribed by (\ref{A8}), where $\;\Phi\;=\;{\partial \, \Delta
{\cal H}}/{\partial p}\;$. This is equivalent to our Theorem from
subsection 2.3. Evidently, for disturbances dependent solely upon
the coordinates, we return to the case explained in subsection 3.2
(equations (\ref{55} - \ref{56})): in that case, the Hamiltonian
formulation of the problem demanded imposition of the Lagrange
constraint (\ref{56}).

To draw to a close, the generalised Lagrange constraint,
$\;\Phibold\,=\,-\;{\partial \,\Delta {\cal L}}/{\partial \dot
q}\;$, is stiffly embedded in the Hamilton-Jacobi technique. Hence
this technique breaks the gauge invariance and is unfit (at least,
in its straightforward form) to describe the gauge symmetry of the
planetary equations. It is necessary to sacrifice gauge freedom by
imposing the generalised Lagrange constraint to make use of the
Hamilton-Jacobi development.

In this special gauge, the perturbed momentum coincides with the
unperturbed one (which was equal to $\;\bf\vec g\;$). Indeed, we
can rewrite (\ref{5559}) as
 \ba
 p\,\equiv\,\frac{\partial\,\left(\,
{\cal L}(q,\,\dot{q},\,t)\,+\,\Delta {\cal
L}(q,\,\dot{q},\,t)\,\right)}{\partial \dot
q}\;=\;\dot{q}\;-\;\Phi\;=\;g\;\;
 \label{5555}
 \ea
which means that, in the astronomical implementation of this
theory, the Hamilton-Jacobi treatment necessarily demands the
orbital elements to osculate in the phase space. Naturally, this
demand reduces to that of regular osculation in the simple case of
velocity-independent $\;\Delta {\cal L}\;$ that was explored in
subsection 3.2.

% \pagebreak

\section{Conclusions}

We have studied, in an arbitrary gauge, the VOP method in
celestial mechanics in the case when the perturbation depends on
both positions and velocities. Such situations emerge when
relativistic corrections to the Newton law are taken into account
or when the VOP method is employed in noninertial frames of
reference (a satellite orbiting a precessing planet being one such
example). The gauge-invariant (and generalised to the case of
velocity-dependent disturbances) Delaunay-type system of equations
is not canonical. We, though, have proven a theorem establishing a
particular gauge (which coincides with the Lagrange gauge in the
absence of velocity dependence of the perturbation) that renders
this system canonical. We called that gauge the "generalised
Lagrange gauge."

We have explained where the Lagrange constraint tacitly enters the
Hamilton-Jacobi derivation of the Delaunay equations. This
constraint turns out to be an inseparable (though not easily
visible) part of the method: in the case of momentum-independent
disturbances, the N-body generalisation of the two-body
Hamilton-Jacobi technique is legitimate only if we use orbital
elements that are osculating, i.e., if we exploit only the
instantaneous ellipses (or hyperbolae, in the flyby case) that are
always tangential to the velocity vector. Oddly enough, an
explicit mention of this circumstance has not appeared in the
astronomical literature (at least, to the best of our knowledge).

In the case of momentum-dependent disturbances, the above
restriction generalises, in that the instantaneous ellipses
(hyperbolae) must be osculating in the phase space. This is
equivalent to the imposition of the generalised Lagrange gauge.

Comparing the good old VOP method with that based on the Jacobi
theorem, we have to acknowledge that the elegance of the latter
does not outweigh the power of the former. If we decide to explore
the infinite multitude of gauges or to study the
numerical-error-invoked gauge drift, we shall not be able to
employ the Hamilton-Jacobi theory without additional structure.
However, the direct VOP method unencumbered with the canonicity
demand will immediately yield gauge-invariant equations for the
Delaunay elements obeying an arbitrary gauge condition
 \be
\sum_i \;\frac{\partial {\bf \vec f}}{\partial D_i}\;\frac{d
D_i}{d t}\;=\; {\bf {\vec \Phi}}(D_i,\,t)\;\;\;,\;
 \label{57}
  \ee
$\Phibold\;$ being some function of time and elements $\;D_i\;$.
In Efroimsky (2002) these equations were written down for the case
of velocity-independent perturbation. If the disturbing force
depends also upon velocities, the Delaunay-type equations will
acquire even more terms and will read as (\ref{Delaunay.1} -
\ref{Delaunay.6}). In the simple case of a velocity-independent
disturbance, any supplementary condition different from that of
Lagrange will drive the Delaunay system away from its canonical
form. If we permit the disturbing force to depend also upon
velocities, the Delaunay equations will retain their canonicity
only in the generalised Lagrange gauge.

In the language of modern physics this may be put in the following
wording. N-body celestial mechanics is a gauge theory but is not
genuinely symplectic insofar as the language of orbital elements
is used. It, though, becomes canonical in the generalised Lagrange
gauge.

The applications of this formalism to motions in non-inertial
frames of reference will be studied in Efroimsky \& Goldreich
(2003). Some other applications were addressed in Slabinski
(2003).

~\\

% ~\\

{\bf Acknowledgments}\\
~\\
The authors are grateful to Jean Kovalevsky for an e-mail
conversation, which he had with one of us (ME) and which initiated
this work. ME would like to acknowledge his numerous stimulating
discussions with Marc Murison, William Newman, and Victor
Slabinski. Research by ME was supported by NASA grant W-19948.
Research by PG was partially supported by NSF grant AST 00-98301.

~\\

% \pagebreak

~\\
{\underline{\bf{\Large{Appendix 1:}}}}\\
~\\ {{{\it{\Large{\bf Gauge-invariant equations of Lagrange and
Delaunay types
\\}}}}}

\noindent We present the gauge-invariant Lagrange-type equations.
They follow  from (\ref{221}) if we take into account the
gauge-invariance of matrix $\,[C_i\,C_j]\,$ defined by (\ref{15}).
We denote by $\Delta {\cal H}$ the perturbation of the
Hamiltonian, connected through (\ref{103}) with that of the
Lagrangian. The latter, in its turn, is connected through
(\ref{105}) with the disturbing force (and acts as the customary
disturbing function when the perturbations are devoid of velocity
dependence).
 \ba
 \nonumber
  \frac{da}{dt}\;=\;\frac{2}{n\,a}\;\;\left[\frac{\partial
 \left(\,-\,\Delta {\cal H} \right)}{\partial M_o} \;-\;\frac{\partial \,\Delta {\cal L}}{\partial {\bf
{\dot {\bf \vec r}}}^{
 \left.~\right.} } \,
 \frac{\partial }{\partial M_o}
 \left({\bf {\vec \Phi}}\,+\,\frac{
 \partial {\,\Delta {\cal L} }}{\partial \bf\dot{\vec r}}\right)
 \;-\right.~~~~~~~~~~~~~~~~~~\\
 \label{222}\\
 \nonumber
\left. \left({\bf{\vec \Phi}}\,+\,\frac{\partial \,\Delta {\cal
L}}{\partial \bf\dot{\vec r}}\right)\,\frac{\partial \bf {\vec
g}}{\partial M_o}
  \;-\;\frac{\partial{\bf{\vec f}}}{\partial
M_o}\, \;\frac{d}{dt} \left({\bf {\vec \Phi}}\,+\,\frac{\partial
\,\Delta {\cal L}}{\partial \bf\dot{\vec r}}\right) \right]\;\;\;,
 \ea
 \ba
 \nonumber\\
 \nonumber
 \frac{de}{dt}\,=\,\frac{1-e^2}{n\,a^2\,e}\;\left[\frac{\partial
 \left(\,-\,\Delta {\cal H} \right)}{\partial M_o} \;-\;\frac{\partial \,\Delta {\cal L}}{\partial {\bf {\dot
{{\bf\vec r}}}}^{
 \left.~\right.} } \,\frac{\partial }{\partial a}\left({\bf
 {\vec \Phi}}\,+\,\frac{\partial {\,\Delta {\cal L}}}{\partial
 \bf\dot{\vec r}}\right) \,-\;~~~~~~~~~~~~~~~~~~~~~~~~~~~~~~\right.\\
 \nonumber\\
 \nonumber\\
 \nonumber
 \left.
 \left({\bf {\vec
\Phi}}\,+\,\frac{\partial \,\Delta {\cal L}}{\partial \bf\dot{\vec
r}}\right)\,\frac{\partial \bf {\vec g}}{\partial M_o} \; -
\frac{\partial{\bf{\vec f}}}{\partial M_o} \,\;\frac{d}{dt}
\left({\bf {\vec \Phi}}\,+\,
 \frac{\partial {\,\Delta {\cal L}}}{\partial \bf\dot{\vec r}}\right)
\right]\;-\;\;\;\;\;\;\;\;\;\;\;\;\;\;\;\;\;\;\;\\
 \label{223}\\
 \nonumber\\
 \nonumber
\frac{(1\,-\,e^2)^{1/2}}{n\,a^2\,e} \;\left[\frac{\partial
 \left(\,-\,\Delta {\cal H} \right)}{\partial \omega} \;-\;
 \frac{\partial \,\Delta {\cal L}}{\partial {\bf {\dot
{{\bf\vec r}}}}^{
 \left.~\right.} }
  \,\frac{\partial }{\partial \omega}\left({\bf {\vec \Phi}}\,+
  \,\frac{\partial \,\Delta {\cal L}}{\partial \bf\dot{\vec r}}\right)
   \;-~~~~~~~~\right.\\
 \nonumber\\
 \nonumber\\
 \nonumber
 \left.
 \left({\bf {\vec
\Phi}}\,+\,\frac{\partial \,\Delta {\cal L}}{\partial \bf\dot{\vec
r}}\right)\,\frac{\partial \bf {\vec g}}{\partial
\omega}\;-\;\frac{\partial{\bf \vec f}}{\partial \omega}
\,\;\frac{d}{dt} \left({\bf {\vec \Phi}}\,+\,\frac{\partial
\,\Delta {\cal L}}{\partial \bf\dot{\vec r}}\right)
\right]\;\;\;\;,\;\;\;\;
 \ea
 ~\\
 \ba
 \nonumber
 \frac{d\omega}{dt}\;=\;\frac{\;-\;\cos \inc
}{n\,a^2\,(1\,-\,e^2)^{1/2}\, \sin \inc }\;\;\left[\frac{\partial
 \left(\,-\,\Delta {\cal H} \right)}{\partial \inc }
  \;-\;\frac{\partial \,\Delta {\cal L}}{\partial {\bf {\dot
{{\bf\vec r}}}}^{
 \left.~\right.} }
  \,\frac{\partial }{\partial \inc}\left({\bf {\vec \Phi}}\,
  +\,\frac{\partial \,\Delta {\cal L}}{\partial \bf\dot{\vec r}}\right)
   \;-\;~~~~~~~~\right.\\
 \nonumber\\
 \nonumber\\
 \nonumber
 \left.
   \left({\bf {\vec \Phi}}\,+\,\frac{\partial
\,\Delta {\cal L}}{\partial \bf\dot{\vec
r}}\right)\,\frac{\partial \bf {\vec g}}{\partial \inc }
\;-\;\frac{\partial{\bf{\vec f}}}{\partial \inc }\, \;\frac{d}{dt}
\left({\bf {\vec \Phi}}\,+\,\frac{\partial \,\Delta {\cal
L}}{\partial \bf\dot{\vec r}}\right)
 \right]\;+\;\;\;\\
 \label{224}\\
 \nonumber\\
 \nonumber
 \frac{(1-e^2)^{1/2}}{n\,a^2\,e}\;
  \left[
  \frac{\partial  \left(\,-\,\Delta {\cal H} \right)}{\partial e} \;-\;
  \frac{\partial \,\Delta {\cal L}}{\partial {\bf {\dot {{\bf\vec r}}}}^{
 \left.~\right.} } \,
 \frac{\partial }{\partial e}\left({\bf {\vec \Phi}}\,+\,
 \frac{\partial \,\Delta {\cal L}}{\partial \bf\dot{\vec r}}\right)
  \;-\;~~~~~~~~\right.\\
 \nonumber\\
 \nonumber\\
 \nonumber
 \left.
  \left({\bf {\vec \Phi}}\,+\,\frac{\partial
\,\Delta {\cal L}}{\partial \bf\dot{\vec
r}}\right)\,\frac{\partial \bf {\vec g}}{\partial e}
\;-\;\frac{\partial{\bf{\vec f}}}{\partial e}\, \;\frac{d}{dt}
\left({\bf {\vec \Phi}}\,+\,\frac{\partial \,\Delta {\cal
L}}{\partial \bf\dot{\vec r}}\right)
 \right]\;\;\;,\;\;\;\;\;\;\;\;\;
 \ea
 ~\\
 \ba
 \nonumber
 \frac{d \inc }{dt}\;=\;\frac{\cos
\inc}{n\,a^2\,(1\,-\,e^2)^{1/2}\, \sin
\inc}\;\;\left[\frac{\partial  \left(\,-\,\Delta {\cal H}
\right)}{\partial \omega} \;-\;\frac{\partial \,\Delta {\cal
L}}{\partial {\bf {\dot {{\bf\vec r}}}}^{
 \left.~\right.} } \,\frac{\partial }{\partial \omega}
 \left({\bf {\vec \Phi}}\,+\,\frac{\partial \,\Delta {\cal L}}{\partial
 \bf\dot{\vec r}}\right) \;-~~~~~~~~\right.\\
 \nonumber\\
 \nonumber\\
 \nonumber
 \left.
 \left({\bf {\vec \Phi}}\,+\,\frac{\partial
\,\Delta {\cal L}}{\partial \bf\dot{\vec
r}}\right)\,\frac{\partial \bf {\vec g}}{\partial \omega }
\;-\;\frac{\partial{\bf{\vec f}}}{\partial \omega }\,
\;\frac{d}{dt} \left({\bf {\vec \Phi}}\,+\,\frac{\partial \,\Delta
{\cal L}}{\partial \bf\dot{\vec
r}}\right) \right]\;-\;\;\;\;\;\;\;\;\;\;\;\\
 \nonumber\\
 \label{225}\\
 \nonumber
 \;\frac{1}{n\,a^2\,(1\,-\,e^2)^{1/2}\,\sin \inc
}\;\;\left[\frac{\partial  \left(\,-\,\Delta {\cal H}
\right)}{\partial \Omega} \;-\;\frac{\partial \,\Delta {\cal
L}}{\partial {\bf {\dot {{\bf\vec r}}}}^{
 \left.~\right.} } \,\frac{\partial }{\partial \Omega}\left({\bf
  {\vec \Phi}}\,+\,\frac{\partial \,\Delta {\cal L}}{\partial
   \bf\dot{\vec r}}\right) \;-~~~~~~~~\right.\\
 \nonumber\\
 \nonumber\\
 \nonumber
 \left.
 \left({\bf {\vec
\Phi}}\,+\,\frac{\partial \,\Delta {\cal L}}{\partial \bf\dot{\vec
r}}\right)\,\frac{\partial \bf {\vec g}}{\partial \Omega }
\;-\;\frac{\partial{\bf {\vec f}}}{\partial \Omega}\,
\;\frac{d}{dt} \left({\bf {\vec \Phi}}\,+\,\frac{\partial
 \,\Delta {\cal L}}{\partial \bf\dot{\vec r}}\right) \right]\;\;\;,\;\;
 \ea
 ~\\
 \ba
 \nonumber
\frac{d\Omega}{dt}\;=\;\frac{1}{n\,a^2\,(1\,-\,e^2)^{1/2}\,\sin
\inc }\;\; \left[\frac{\partial  \left(\,-\,\Delta {\cal H}
\right)}{\partial \inc } \;-\;\frac{\partial \,\Delta {\cal
L}}{\partial {\bf {\dot {{\bf\vec r}}}}^{
 \left.~\right.} } \,\frac{\partial }{\partial \inc}\left({\bf
  {\vec \Phi}}\,+\,\frac{\partial
   \,\Delta {\cal L}}{\partial \bf\dot{\vec r}}\right) \;-~~~~~~~~\right.\\
 \nonumber\\
 \label{226}\\
 \nonumber
 \left.\left( {\bf {\vec
\Phi}}\,+\,\frac{\partial \,\Delta {\cal L}}{\partial \bf\dot{\vec
r}}\right)\,\frac{\partial \bf {\vec g}}{\partial \inc }
\;-\;\frac{\partial{\bf{\vec f}}}{\partial \inc }\, \;\frac{d}{dt}
\left({\bf {\vec \Phi}}\,+\,\frac{\partial \,\Delta {\cal
L}}{\partial \bf\dot{\vec r}}\right)
 \right]\;\;\;,\;\;\;\;\;\;\;\;\;
 \ea
 ~\\
 \ba
 \nonumber
\frac{dM_o}{dt}\,=\,\;-\,\frac{1\,-\,e^2}{n\,a^2\,e}\,\;\left[
\frac{\partial  \left(\,-\,\Delta {\cal H} \right)}{\partial e }
\;-\;\frac{\partial \,\Delta {\cal L}}{\partial {\bf {\dot
{{\bf\vec r}}}}^{ \left.~\right.} } \,\frac{\partial }{\partial
e}\left({\bf {\vec \Phi}}\,+\,\frac{\partial \,\Delta {\cal
L}}{\partial \bf\dot{\vec
r}}\right) \,-~~~~~~~~\right.\\
 \nonumber\\
 \nonumber\\
 \nonumber
 \left.\left({\bf {\vec \Phi}}\,+\,\frac{\partial \,\Delta {\cal L}}{\partial \bf\dot{\vec r}}\right)\,\frac{\partial \bf {\vec
g}}{\partial e} \,-\,\frac{\partial{\bf{\vec f}}}{\partial e }
\,\;\frac{d}{dt} \left({\bf {\vec \Phi}}\,+\,\frac{\partial
\,\Delta {\cal L}}{\partial \bf\dot{\vec r}}\right)
 \right]\;-\\
 \label{227}\\
 \nonumber\\
 \nonumber
 \frac{2}{n\,a}\,\left[\frac{\partial  \left(\,-\,\Delta {\cal H} \right)}{\partial a }
\;-\;\frac{\partial \,\Delta {\cal L}}{\partial {\bf {\dot
{{\bf\vec r}}}}^{
 \left.~\right.} } \,\frac{\partial }{\partial a}
 \left({\bf {\vec \Phi}}\,+\,\frac{\partial \,\Delta {\cal L}}{\partial
  \bf\dot{\vec r}}\right) \,-~~~~~~~~\right.\\
 \nonumber\\
 \nonumber\\
 \nonumber
 \left. \left({\bf {\vec
\Phi}}\,+\,\frac{\partial \,\Delta {\cal L}}{\partial \bf\dot{\vec
r}}\right)\,\frac{\partial \bf {\vec g}}{\partial a}
\,-\,\frac{\partial{\bf{\vec f}}}{\partial a } \,\;\frac{d}{dt}
\left({\bf {\vec \Phi}}\,+\,\frac{\partial \,\Delta {\cal
L}}{\partial \bf\dot{\vec r}}\right)
 \right]\;\;\;\;.
 \ea
Similarly, the gauge-invariant Delaunay-type system can be written
down as:
 \ba
\frac{dL}{dt}=\frac{\partial  \left(\,-\,\Delta {\cal
H}\right)}{\partial M_o}-\frac{\partial \,\Delta  L}{\partial
{\bf{\dot{\vec{r}}}} }\,\frac{\partial }{\partial M_o}\left({\bf
{\vec \Phi}}\,+\,\frac{\partial \,\Delta {\cal L}}{\partial
\bf\dot{\vec r}}\right) -\left({{\bf\vec\Phi}}\,+\,\frac{\partial
\,\Delta {\cal L}}{\partial \bf\dot{\vec r}}\right)\,
\frac{\partial {\bf \vec g}}{\partial M_o}-\frac{\partial {\bf
\vec f}}{\partial M_o}\,\frac{d}{dt} \left({\bf {\vec
\Phi}}\,+\,\frac{\partial \,\Delta {\cal L}}{\partial \bf\dot{\vec
r}}\right)\;\;,\;\;
 \label{Delaunay.1}\\
 \nonumber\\
\frac{d M_o}{dt}\;=\,-\,\frac{\partial \left(\,-\,\Delta {\cal
H}\right)}{\partial L}+\frac{\partial \,\Delta {\cal L}}{\partial
{\bf{\dot{\vec{r}}}} }\,\frac{\partial }{\partial L}\left({\bf
{\vec \Phi}}\,+\,\frac{\partial \,\Delta {\cal L}}{\partial
\bf\dot{\vec r}}\right)+\left({{\bf\vec\Phi}}\,+\,\frac{\partial
\,\Delta {\cal L}}{\partial \bf\dot{\vec r}}\right)
\,\frac{\partial {\bf \vec g}}{\partial L}+\frac{\partial {\bf
\vec f}}{\partial L}\,\frac{d}{dt} \left({\bf {\vec
\Phi}}\,+\,\frac{\partial \,\Delta {\cal L}}{\partial \bf\dot{\vec
r}}\right)\;\;\;\;,\;\;\;
 \label{Delaunay.2}
 \ea
 \ba
 \frac{dG}{dt}\;=\;\frac{\partial \left(\,-\,\Delta {\cal H}\right)}{\partial
\omega}\;-\;\frac{\partial \,\Delta {\cal L}}{\partial
{\bf{\dot{\vec{r}}}} }\,\frac{\partial }{\partial
\omega}\left({\bf {\vec \Phi}}\,+\,\frac{\partial \,\Delta {\cal
L}}{\partial \bf\dot{\vec r}}\right)\;-\;\left({{\bf\vec
\Phi}}\,+\,\frac{\partial \,\Delta {\cal L}}{\partial \bf\dot{\vec
r}}\right)\;\frac{\partial {\bf \vec g}}{\partial
\omega}\;-\;\frac{\partial {\bf \vec f}}{\partial
\omega}\;\frac{d}{dt} \left({\bf {\vec \Phi}}\,+\,\frac{\partial
\,\Delta {\cal L}}{\partial \bf\dot{\vec r}}\right)\;\;\;,\;\;
 \label{Delaunay.3}\\
 \nonumber\\
\frac{d\omega }{dt}\;=\,-\,\frac{\partial \left(\,-\,\Delta {\cal
H}\right)}{\partial G}\,+\,\frac{\partial \,\Delta {\cal
L}}{\partial {\bf{\dot{\vec{r}}}} }\frac{\partial }{\partial
G}\left({\bf {\vec \Phi}}\,+\,\frac{\partial \,\Delta {\cal
L}}{\partial \bf\dot{\vec
r}}\right)\,+\,\left({{\bf\vec\Phi}}\,+\,\frac{\partial \,\Delta
{\cal L}}{\partial \bf\dot{\vec r}}\right) \frac{\partial {\bf
\vec g} }{\partial G}\, +\,\frac{\partial {\bf \vec f}}{\partial
G}\,\frac{d}{dt} \left({\bf {\vec \Phi}}\,+\,\frac{\partial
\,\Delta {\cal L}}{\partial \bf\dot{\vec r}}\right)\;\;\;,\;\;
 \label{Delaunay.4}
 \ea
 \ba
 \frac{d H}{dt}\,=\,\frac{\partial \left(\,-\,\Delta {\cal H}\right)}{\partial
\Omega}\,-\,\frac{\partial \,\Delta {\cal L}}{\partial
{\bf{\dot{\vec{r}}}} }\,\frac{\partial }{\partial
\Omega}\left({\bf {\vec \Phi}}\,+\,\frac{\partial \,\Delta {\cal
L}}{\partial \bf\dot{\vec r}}\right)\,-\,\left({\bf\vec
\Phi}\,+\,\frac{\partial \,\Delta {\cal L}}{\partial \bf\dot{\vec
r}}\right)\,\frac{\partial {\bf \vec g}}{\partial \Omega
}\,-\,\frac{\partial {\bf\vec f}}{\partial \Omega }\,\frac{d}{dt}
\left({\bf {\vec \Phi}}\,+\,\frac{\partial \,\Delta {\cal
L}}{\partial \bf\dot{\vec r}}\right)\;\;\;\;,\;\;\;
  \label{Delaunay.5}\\
 \nonumber\\
\frac{d \Omega}{dt}\,=\,-\,\frac{\partial \left(\,-\,\Delta {\cal
H}\right)}{\partial H}\,+\,\frac{\partial \,\Delta {\cal
L}}{\partial {\bf{\dot{\vec{r}}}} }\,\frac{\partial }{\partial
H}\left({\bf {\vec \Phi}}\,+\,\frac{\partial \,\Delta {\cal
L}}{\partial \bf\dot{\vec r}}\right)
 \,+\,\left({\bf\vec\Phi}\,+\,\frac{\partial \,\Delta {\cal L}}{\partial
\bf\dot{\vec r}}\right)\,\frac{\partial {\bf\vec g}}{\partial
H}\,+\,\frac{\partial {\bf \vec f}}{\partial H}\,\frac{d}{dt}
\left({\bf {\vec \Phi}}\,+\,\frac{\partial \,\Delta {\cal
L}}{\partial \bf\dot{\vec r}}\right)\;\;.\;\;
 \label{Delaunay.6}
 \ea
where
 \ba L\,\equiv\,\mu^{1/2}\,a^{1/2}\,\;\;,\;\;\;\;\;
G\,\equiv\,\mu^{1/2}\,a^{1/2}\,\left(1\,-\,e^2\right)^{1/2}\,\;\;,\;\;\;\;
H\,\equiv\,\mu^{1/2}\,a^{1/2}\,\left(1\,-\,e^2\right)^{1/2}\,\cos
\inc\,\;\;\;
 \label{229}
 \ea
and the symbols $\;{\Phibold},\,{\bf{\vec f}},\,{\bf{\vec g}}\;$
denote the functional dependencies of the gauge, position and
velocity upon the Delaunay, not Keplerian elements, and therefore
these are functions different from $\;{\Phibold},\,{\bf{\vec
f}},\,{\bf{\vec g}}\;$ used in (\ref{222} - \ref{227}) where they
stood for the dependencies upon the Kepler elements. (In Efroimsky
(2002) the dependencies $\;{\Phibold},\,{\bf{\vec f}},\,{\bf{\vec
g}}\;$ upon the Delaunay variables were equipped with tilde, to
distinguish them from the dependencies upon the Kepler
coordinates.)

The above equations do not merely repeat those derived earlier in
Efroimsky (2002, 2003), but generalise them to the case of a
perturbation $\;\Delta {\cal L}\;$ which is both position- and
velocity-dependent. For this reason, our gauge-invariant equations
can be employed not only in an inertial frame but also in a
wobbling one.

To employ the gauge-invariant equations in analytical calculations
is a delicate task: one should always keep in mind that, in case
$\;\Phibold\;$ is chosen to depend not only upon time but also
upon the "constants" (but not upon their derivatives), the
right-hand sides of these equation will implicitly contain the
first derivatives $\;dC_i/dt\;$, and one will have to move them to
the left-hand sides (much like in the transition from (\ref{217})
to (\ref{general_F})).

% ~\\

% \pagebreak

 ~\\
 {\underline{\bf{\Large{Appendix 2:}}}}\\
  ~\\ {{{\it{\Large{\bf  The Hamilton-Jacobi method in celestial
 mechanics
 \\}}}}}

The Jacobi equation (\ref{30}) is a PDE of the first order, in
(N+1) variables $\,(q_n,\,t)\,$, and its complete integral
$\;W(q,\,Q,\,t)\;$ will depend upon $\,N+1\,$ constants $\,a_n\;$
(Jeffreys and Jeffreys 1972, Courant and Hilbert 1989). One of
these constants, $\;a_{\smallN+1}\,$, will be additive, because
$\;W\;$ enters the above equation only through its derivatives.
Since both Hamiltonians are, too, defined up to some constant
$\,\it f\,$, then the solution to (\ref{30}) must contain that
constant multiplied by the time:
 \ba
 \nonumber
W(q,\,a_1,\,...\,,\,a_{\smallN},\,a_{\smallN+1} ,\,t)\,=\,\tilde
W(q,\,a_1,\,...\,,\,a_{\smallN},\,t)\,-\,( t -
t_o)\;f(a_1,\,...\,,\,a_{\smallN})\\
 \label{31}\\
\nonumber
 =\;\tilde W(q,\,a_1,\,...\,,\,a_{\smallN},\,t)\,-\,t\;f(a_1,\,...\,,\,a_{\smallN})
\;-\;a_{\smallN+1}
 \ea
where the fiducial epoch is connected to the constants through
$\;t_o\,=\,-\,a_{\smallN + 1}/f\;$, and the function $\;\tilde
W\;$ depends upon $\;N\;$ constants only. As the total number of
independent adjustable parameters is $\,N+1\,$, the constant
$\,\it f\,$ cannot be independent but must rather be a function of
$\;a_1,\,...\,,\,a_{\smallN},\,a_{\smallN + 1}\;$. Since we agreed
that the constant $\,a_{\smallN + 1}\,$ is additive and shows
itself nowhere else, it will be sufficient to consider $\,\it f\,$
as a function of the rest $\,N\,$ parameters only. (In principle,
it is technically possible to involve the constant $\;a_{\smallN +
1}\;$, i.e., the reference epoch, into the mutual transformations
between the other constants. However, in the applications that we
shall consider, we shall encounter only equations autonomous in
time, and so there will be no need to treat $\;a_{\smallN + 1}\;$
as a parameter to vary. Hence, in what follows we shall simply
ignore its existence.) The new function $\;\tilde W\;$ obeys the
simplified Jacobi equation
 \be
{\cal H}\left(q\,,\;\,\frac{\partial \tilde
W(q,\,a_1,\,...\,,\,a_{\smallN},\,t)}{\partial
q}\,,\,t\right)\;+\;\frac{\partial \tilde W
(q,\,a_1,\,...\,,\,a_{\smallN},\,t)}{\partial t}\;
=\;f(a_1,\,...\,,\,a_{\smallN})\;+\;{\cal H}^{*}
 \label{32}
 \ee
As agreed above, $\;{\cal H}^{*}\;$ is a constant. Hence, we can
state about this constant all the same as about the constant
$\,\it f\,$: since the integral $\;W\;$ can contain no more than
$\;N\,+\,1\;$ adjustable parameters
$\;a_1,\,...\,,\,a_{\smallN},\,a_{\smallN+1}\;$, and since we
ignore the existence of $\,a_{\smallN + 1}\,$, the constant
$\;{\cal H}^{*}\;$ must be a function of the remaining N
parameters: $\;{\cal H}^{*}\;=\;{\cal
H}^{*}(a_1,\,...\,,\,a_{\smallN})$.

Now, in case $\;\cal H\;$ depends only upon $\;(q,\,p)\;$ and
lacks an explicit time dependence, then so will $\;\tilde W$; and
the above equation will very considerably simplify:
 \be
{\cal H}\left(q\,,\;\,\frac{\partial \tilde
W(q,\,a_1,\,...\,,\,a_{\smallN})}{\partial q}\right)\;
=\;f(a_1,\,...\,,\,a_{\smallN})\;+\;{\cal
H}^{*}(a_1,\,...\,,\,a_{\smallN})\;\;\;,
 \label{33}
 \ee
where we deliberately avoided absorbing the constant Hamiltonian
$\;{\cal H}^{*}\;$ into the function $\,\it f\,$.

Whenever the integral $\;W\;$ can be found explicitly, the
constants $\;(a_1, \,...\,,\,a_{\smallN})\;$ can be identified
with the new coordinates $\,Q\,$, whereafter the new momenta will
be calculated through $\;P\,=\,-\,\partial W / \partial Q\;$. In
the special case of zero $\;{\cal H}^{*}\;$, the new momenta
become constants, because they obey the canonical equations with a
vanishing Hamiltonian. In the case where $\;{\cal H}^{*}\;$ is a
nonzero constant, it must, as explained above, be a function of
all or some of the independent parameters $\;(a_1,
\,...\,,\,a_{\smallN})\;$, and therefore, all or some of the new
momenta $\,P\,$ will be evolving in time.

Since it is sufficient to find only one solution to the Jacobi
equation, one can seek it by means of the variable-separation
method: equation (\ref{33}) will solve in the special case when
the generating function (\ref{31}) is separable:
 \be
\tilde
W(q_1,\,...\,,\,q_{\smallN},\,a_1,\,...\,,\,a_{\smallN})\;=\;\sum_{i=1}^N
\tilde W_i \left(q_i\,,\, a_1,\,...\,,\,a_{\smallN}  \right)
 \label{34}
 \ee
This theory works very well in application to the unperturbed
(two-body) problem (\ref{1}) of celestial mechanics, a problem
that is simple due to its mathematical equivalence to the
gravitationally bound motion of a reduced mass
$\;\;m_{planet}m_{sun}/(m_{planet} + m_{sun})\;\;$ about a fixed
centre of mass $\;\;m_{planet} + m_{sun}\;$. If one begins with
the (reduced) two-body Hamiltonian in the spherical coordinates
 \be
q_1\;=\;r\;\;\;,\;\;\;\;\;\;q_2\;=\;\phi\;\;\;,\;\;\;\;\;\;q_3\;=\;\theta
 \label{35}
 \ee
 (where $x=r\,\cos\phi\;\cos\theta\;,\;\;
 y=r\,\cos\phi\;\sin\theta\;,\;\;z=r\,\sin\phi $),
then the expression for Lagrangian,
 \ba
 L\;=\;T\;-\;\Pi\;=\;\frac{1}{2}\,\left(\dot{q}_1\right)^2\;+\;
 \frac{1}{2}\,\left({q}_1\right)^2\;\left(\dot{q}_2\right)^2\;+\;
 \frac{1}{2}\,\left({q}_1\right)^2\;\left(\dot{q}_3\right)^2\;\cos^2q_2\;+\;
\frac{\mu}{q_1}
 \label{36}
 \ea
will yield the following formulae for the momenta:
 \ba
 p_1\;\equiv\;\frac{\partial \cal L}{\partial \dot{q}_1}\;=\;\dot{q}_1 \;\;\;\;,
\;\;\;\;\;\;
 p_2\;\equiv\;\frac{\partial \cal L}{\partial
\dot{q}_2}\;=\;q_1^2\,\dot{q}_2
 \;\;\;\;,\;\;\;\;\;\;p_3\;\equiv\;\frac{\partial \cal L}{\partial
 \dot{q}_3}\;=\;q_1^2\,\dot{q}_3\;\cos^2q_2
 \label{37}
 \ea
whence the initial Hamiltonian will read:
 \ba
 {\cal H}\;=\;\sum p
 \,\dot{q}\;-\;{\cal L}\;=\;\frac{1}{2}\,p_1^2\;+\;\frac{1}{2q_1^2}\,p_2^2\;+\;
 \frac{1}{2q_1^2\cos^2q_2}\,p_3^2
 \;-\;\frac{\mu}{q_1}\;\;\;.
 \label{38}
 \ea
Then the Hamilton-Jacobi equation (30) will look like this:
 \ba
\frac{1}{2}\,\left( \frac{\partial W}{\partial q_1 }
\right)^2\;+\;\frac{1}{2q_1^2}\,\left( \frac{\partial W}{\partial
q_2 } \right)^2\;+\;
 \frac{1}{2q_1^2\cos^2q_2}\,\left( \frac{\partial W}{\partial q_3 }
\right)^2
 \;-\;\frac{\mu}{q_1}\;-\;\frac{\partial W}{\partial
 t}\;-\;{\cal H}^{*}\;=\;0\;\;,
\label{39}
 \ea
 while the auxiliary function $\;\tilde W\;$ defined through
 (\ref{31}) will obey
 \ba
\frac{1}{2}\,\left( \frac{\partial \tilde W}{\partial q_1 }
\right)^2+\;\frac{1}{2q_1^2}\,\left( \frac{\partial \tilde
W}{\partial q_2 } \right)^2+\;
 \frac{1}{2q_1^2\cos^2q_2}\,\left( \frac{\partial \tilde W}{\partial q_3 }
\right)^2
 -\;\frac{\mu}{q_1}\;-\;f\;-\;{\cal H}^{*}\,=\;0\;\;.\;\;\;
 \label{40}
 \ea
A lengthy but elementary calculation (presented, with some
inessential variations, in Plummer 1918, Smart 1953, Pollard 1966,
Kovalevsky 1967, Stiefel and Scheifele 1971, and many other books)
shows that, for a constant $\;{\cal H}^{*}\;$ and in the ansatz
(\ref{34}), the integral of (\ref{33}) takes the form:
 \ba
 \nonumber
\tilde W\;=\;\tilde W_1\left(q_1,\,a_1,\,a_2,\,a_3\right)\;+\;
\tilde W_2\left(q_2,\,a_1,\,a_2,\,a_3\right)\;+\;
\tilde W_3\left(q_3,\,a_1,\,a_2,\,a_3\right)\;=\;\;\;\;\;\;\;\;\;\;\;\;\;\;\;\\
 \label{41}\\
 \nonumber
 \int^{q_1(t)}_{q_1(t_o)} \;\epsilon_1\;\left(2\left(f\,+\,{\cal H}^{*}\right)\;+\;
\frac{2\mu}{q_1}\;-\;\frac{a_2^2}{q_1^2}
 \right)^{1/2}d q_1 \; +
\;\int^{\phi}_{0} \;\epsilon_2\; \left(
a_2^2\;-\;\frac{a_3^2}{\cos^2 q_2} \right)^{1/2}dq_2\;+\;
\int_{0}^{\theta} a_3 \,\;dq_3
 \;
 \ea
where the epoch and factors $\epsilon_{1},\,\epsilon_2$ may be
taken as in Kovalevsky (1967): time $t_o$ is the instant of
perigee passage;  factor $\epsilon_{1}$ is chosen to be $+\,1$
when $q_1\equiv\,r$ is increasing, and is $-\,1$ when $r$ is
decreasing;  factor $\epsilon_{2}$ is $+\,1$ when
$q_2\equiv\,\phi$ is increasing, and is $-\,1\,$ otherwise. This
way the quantities under the first and second integration signs
have continuous derivatives. To draw conclusions, in the two-body
case we have a transformation-generating function
 \ba
 \nonumber
W\;\equiv\;\tilde
W\;+\;t\;f(a_1,\,...\,,\,a_{\smallN})\;=\;\;\;\;\;\;\;\;\;\;\;\;\;\;\;
 \;\;\;\;\;\;\;\;\;\;\;\;\;\;\;\\
 \label{42}\\
 \nonumber
 \int^{q_1(t)}_{q_1(t_o)}
\,\epsilon_1\,\left(2\left(f\,+\,{\cal H}^{*}
 \right)\,+\,\frac{2\mu}{q_1}\,-\,\frac{a_2^2}{q_1^2}
 \right)^{1/2}d q_1 \, +
\,\int^{\phi}_{0} \,\epsilon_2\, \left(
a_2^2\,-\,\frac{a_3^2}{\cos^2 q_2} \right)^{1/2}dq_2
 \,+\, \int_{0}^{\theta} a_3 \,\;dq_3\,+\,t\,f
 \ea
whose time-independent component $\;\tilde W\;$ enters equation
(\ref{33}). The first integration in (\ref{42}) contains the
functions $\;f(a_1,\,...\,,\,a_{\smallN})\;$ and $\;{\cal
H}^{*}(a_1,\,...\,,\,a_{\smallN})\;$, so that in the end of the
day $\;W\;$ depends on the N constants
$\;a_1,\,...\,,\,a_{\smallN}\;$ (not to mention the neglected
$\;t_o\;$, i.e., the $\;a_{\smallN + 1}\;$).

Different authors deal differently with the sum $\,(f\,+\,{\cal
H}^{*})\,$ emerging in (\ref{42}). Smart (1953) and Kovalevsky
(1967) prefer to put
 \ba
 f\,=\,0\;\;\;, \;\;\;\;\;\;\;{\cal H}^{*}\,=\,a_1\,\;\;\;,\;\;\;\;\;\;\;a_1\,
=\,-\mu/(2a)\;\;\;,
 \label{43}
 \ea
whereupon the new momentum $\;P_1\,=\,-\,\partial W/\partial
Q_1\,=\,-\,\partial W/\partial a_1\;$ becomes time-dependent (and
turns out to equal $\;-\,t\,+\,t_o\;$.) An alternative choice,
which, in our opinion, better reflects the advantages of the
Hamilton-Jacobi theory, is furnished by Plummer (1918):
 \ba
 f\,=\,a_1\;\;\;, \;\;\;\;\;\;\;{\cal H}^{*}\,=\,0\,\;\;\;,\;\;\;\;\;\;\;a_1\,=\,
 \sqrt{\mu
 \,a}\;\;\;.
 \label{44}
 \ea
 This entails the following correspondence between the new canonical variables
 (the Delaunay elements) and the Keplerian orbital coordinates:
 \ba
 \nonumber
 Q_1\;\equiv\;a_1\;=\;\sqrt{\mu\;a}\;\;\;\;\;\;\;\;\;\;\;\;\;\;\;\;\;\;\;\;\;\;
 \;\;\;,\;\;\;\;\;\;\;\;
 P_1\;=\;-\;M_o\;\;\;\;\;\;\;\;\;\;\;\;\;\;\;,
 \\
 \nonumber\\
 Q_2\;\equiv\;a_2\;=\;\sqrt{\mu\,a\,\left(1\,-\,e^2\right)}\;\;\;\;\;\;\;\;\;\;
 \;\;,\;\;\;\;\;\;\;
 P_2\;=\;-\;\omega\;\;\;\;\;\;\;\;\;\;\;\;\;\;\;\;\;\;,
 \label{45}\\
 \nonumber\\
 \nonumber
 Q_3\;\equiv\;a_3\;=\;\sqrt{\mu\,a\,\left(1\,-\,e^2\right)}\,\cos\,i\;\;\;\;,\;
 \;\;\;\;\;\;
 P_3\;=\;-\;\Omega\;\;\;\;\;\;\;\;\;\;\;\;\;\;\;\;\;\;.
 \ea
 Everywhere in this paper we follow the convention (\ref{44}) and
 denote
 the above variables $Q_1, \,Q_2,\,Q_3$ by $L,\,G,\,H$,
 correspondingly (as is normally done in the astronomical literature)

~\\

% \pagebreak

\end{document}